\documentclass[journal,symmetry,article,submit,moreauthors,pdftex]{mdpi} 

\firstpage{1} 
\pubvolume{1}
\issuenum{1}
\articlenumber{0}
\pubyear{2020}
\copyrightyear{2020}
\history{Received: date; Accepted: date; Published: date}

\Title{Probing the nuclear equation of state from the existence of a $\sim 2.6~M_{\odot}$ neutron star: the GW190814 puzzle}

\Author{Alkiviadis Kanakis-Pegios$^{\dagger,\ddagger}$\orcidA{}, Polychronis S. Koliogiannis $^{*,\dagger,\ddagger}$\orcidB{} and Charalampos C. Moustakidis$^{\dagger,\ddagger}$\orcidC{}}

\AuthorNames{A. Kanakis-Pegios, P.S. Koliogiannis and Ch.C. Moustakidis}

\address[1]{%
Department of Theoretical Physics, Aristotle University of Thessaloniki, 54124 Thessaloniki, Greece; alkanaki@auth.gr (A.K.P.); moustaki@auth.gr (Ch.C.M.)}
\corres{Correspondence: pkoliogi@physics.auth.gr}

\firstnote{Current address: Aristotle University of Thessaloniki, 54124 Thessaloniki, Greece} 
\secondnote{These authors contributed equally to this work.}

\abstract{On August 14, 2019, the LIGO/Virgo collaboration observed a compact object with mass $\sim 2.59_{-0.09}^{+0.08}~M_{\odot}$, as a component of a system where the main companion was a black hole with mass $\sim 23~M_{\odot}$. A scientific debate initiated concerning the identification of the low mass component, as it falls into the neutron star - black hole mass gap. The understanding of the nature of GW190814 event will offer rich information concerning open issues, the speed of sound and the possible phase transition into other degrees of freedom. In the present work, we made an effort to probe the nuclear equation of state along with the GW190814 event. Firstly, we examine possible constraints on the nuclear equation of state inferred from the consideration that the low mass companion is a slow or rapidly rotating  neutron star. In this case, the role of the upper bounds on the speed of sound is revealed, in connection with the dense nuclear matter properties. Secondly, we systematically study the tidal deformability of a possible high mass candidate existing as an individual star or as a component one in a binary neutron star system. As the tidal deformability and radius are quantities very sensitive on the neutron star equation of state, they are excellent counters on dense matter properties. We conjecture that similar isolated neutron stars or systems may exist in the Universe and their possible future observation will shed light on the maximum neutron star mass problem.}

\keyword{Equation of state; Neutron star; GW190814; Maximum mass; Tidal deformability}  

\begin{document}

\section{Introduction}
In Ref.~\cite{Abbott-4} the authors reported the observation of a compact binary coalescence involving a $22.2-24.3\ M_{\odot}$ black hole  and a compact object with a mass of a $2.50-2.67 \ M_{\odot}$ (all measurements quoted at the $ 90 \%$ credible level). The announcement of the GW190814 event~\cite{Abbott-4} triggered various theoretical studies concerning the equation of state (EoS) of dense nuclear matter, in order to explain the possibility of the second partner to be a very massive neutron star (NS). It is worth pointing out that the authors in Ref.~\cite{Abbott-4} did not exclude the possibility that the second partner to be a NS or an exotic compact object, i.e. quark star, boson star or gravastar. 

It is worth to point out the observation of the GW190814 event has some additional general benefits apart from the measurement of 2.6 $M_{\odot}$ of the second partner~\cite{Abbott-4}. Firstly, this binary system  has the most unequal mass ratio yet measured with gravitational waves close to the value of $0.112$. Secondly, the dimensionless of the spin of the primary black hole is constrained $\leq 0.07$, where various tests of general relativity confirm this value, as well as its predictions of higher -multiple emission at high confidence interval. Moreover, the GW190814 event poses a challenge for the understanding of the population of merging compact binaries. It was found after systematic analysis that the merger rate density of GW190814-like binary system to be $7_{-6}^{+16}$ Gpc$^{-3}$yr$^{-1}$~\cite{Abbott-4}. More relevant to the present study, the observation of the GW190814 event led to the following conclusion: due to the source's asymmetric masses, the lack of detection of an electromagnetic counterpart and of clear signature of tides or spin-induced quadrupole effect in the waveform of the gravitational waves we are not able to distinguish between a  black hole-black hole and black hole-neutron star system~\cite{Abbott-4}. In this case, one must count only to the comparison between the mass of the second partner with the estimation of maximum NSs mass $M_{\rm max}$~\cite{Datta-2020}. This is one of the subjects of the present work. It should be emphasized that the measurements of NSs mass can also inform us about a bound on the $M_{\rm max}$  independently of the assumptions of specific EoS. For example, Alsing \textit{et al.}~\cite{Alsing-2018} fitting the known population of NSs in binaries to double-Gaussian mass distribution  obtained the empirical constraint that $M_{\rm max} \leq 2.6 \ M_{\odot}$ (with $90 \%$ confidence interval ). Moreover, Farr and Chatziionannou~\cite{Farr-2020} updated the previous study including recent measurements. Their study constraint the maximum mass $M_{\rm max}=2.25_{-0.26}^{+0.81} \ M_{\odot}$ leading to the conclusion that the posterior probability (for the mass of the second partner $m_2 \leq M_{\rm max}$) is around only $29 \%$. However, the prediction of $M_{\rm max}$ is sensitive on the selection rules mass of NSs (not only on binary systems but also isolated) as well as on the discovery of new events and consequently remains an open problem. Finally, the conclusion of the recent  GW190814  event in comparison with previous ones (for example the GW170817 event~\cite{Abbott-2018}) may shed light on the problem of the $M_{\rm max}$. For example, the spectral EoSs which are conditioned by the  GW170817 event, are once more elaborated to include the possibility that of the prediction of $M_{\rm max}$  at least equal to $m_2$. This approach lead to significant constraints on the radius and tidal deformability  of a NS with mass of $1.4 \ M_{\odot}$ ($R_{1.4}=12.9_{-0.7}^{+0.8}$ km and $\Lambda_{1.4}=616_{-158}^{+273}$ respectively~\cite{Abbott-4}).

The consideration of a NS as the second partner has been studied in recent Refs.~\cite{Tsokaros-20,Huang-2020,Kalogera-2020,Fattoyev-2020,Essick-2020,Safarzadeh-2020,Godzieba-2020,Sedrakian-2020,Biswas-2020,Zhang-2-2020,Most-2020,Tan-2020}. On the other hand, the case that the second partner is a quark or hybrid star has been explored in Refs.~\cite{Zhang-2020,Bombaci-2020,Demircik-2020,Cao-2020,Dexheimer-2020,Roupas-2020,Rather-2020}. Finally, some modified theories of gravity have also been applied as possible solutions to the problem~\cite{Moffat-2020,Oikonomou-2020,Nunes-2020}. In particular, Tsokaros \textit{et al.}~\cite{Tsokaros-20} showed using viable equations that rapid uniform rotation is adequate to explain the existence of a stable $2.6 \ M_{\odot}$ NS for moderately stiff EoSs but may not be adequate for soft ones. Huang \textit{et al.}~\cite{Huang-2020} concluded that using the density-dependent relativistic mean field model one cannot exclude the possibility of the secondary object to be a NS composed of hadronic matter. Zevin \textit{et al.}~\cite{Kalogera-2020} performing a systematic study  led to the conclusion that the formation of GW190814-like systems at any measurable rate requires a supernova engine model that acts on longer timescales such that the proto-compact object can undergo substantial accretion immediately before the explosion. This conclusion hinds that if GW190814 is the result of a massive star binary evolution, the mass gap between NSs and black holes may be narrower or nonexistent. Fattoyev \textit{et al.}~\cite{Fattoyev-2020} speculated that the maximum NS mass can not be significantly higher than the existing observational limit and also the $2.6  M_{\odot}$ compact object is likely to be the lightest black hole ever discovered. Essick and Landry~\cite{Essick-2020} found that there is a $\leq 6 \%$  chance that GW190814 involved a slowly spinning NS, regardless of their assumed population model (considering no overlap between the NS and black hole mass distributions). Safarzadeh and Loeb~\cite{Safarzadeh-2020} suggested that the secondary partner was born as a NS  where a significant amount of the supernova ejecta mass from its formation remained bound to the binary due to the presence of the massive black hole  companion. The bound mass forms a circumbinary accretion disk and its accretion onto the NS created a mass gap object. Godzieba \textit{et al.}~\cite{Godzieba-2020}, showed  how a lower limit on the maximum mass of NSs, in combination with upcoming measurements of NS radii by LIGO/Virgo and NICER, would constrain the EoS of dense matter and  discussed  the implications for the GW190814 event. Sedrakian \textit{et al.}~\cite{Sedrakian-2020}, allowing the hyperonization of dense matter, found that the maximal masses of hypernuclear stars, even for maximally rotating configurations, are inconsistent with a stellar nature interpretation of the light companion in GW190814. They concluded that the GW190814 event involved two black holes rather than a NS and a black hole. In addition, Li \textit{et al.}~\cite{Li-2020} using a set of hadronic EoSs derived from covariant density functional theory to study the impact of their high-density behavior on the properties of rapidly rotating $\Delta$-resonance-admixed hyperonic compact stars, concluded that the low mass companion of GW190814 event is likely to be a low-mass black hole rather than a supramassive neutron star. Biswas \textit{et al.}~\cite{Biswas-2020} concluded that the odds of the secondary object in GW190814 being a NS improved by considering a stiff high-density EoS or a large rotation. Zhang and Li~\cite{Zhang-2-2020} showed that one possible explanation for GW190814's secondary component is a super-fast pulsar spinning faster than 971 Hz. Most \textit{et al.}~\cite{Most-2020} stated that  based on our current understanding of the nuclear matter EoS, it can be a rapidly rotating NS that collapsed to a rotating black hole at some point before the merger. Tan \textit{et al.}~\cite{Tan-2020} constructed heavy NSs by introducing non-trivial structure in the speed of sound sourced by deconfined QCD (Quantum Chromodynamics) matter. Within this approach they can explain the high mass of the second partner. 

Zhang and Mann~\cite{Zhang-2020} indicated a new possibility that the currently observed compact stars, including the recently reported GW190814's secondary component can be quark stars composed of interacting up-down quark matter. Bombaci \textit{et al.}~\cite{Bombaci-2020} investigated the possibility that the low mass companion of the black hole in the source of GW190814 was a strange quark star. This possibility is viable within the so-called two-families scenario in which NSs and strange quark stars coexist. Demircik \textit{et al.}~\cite{Demircik-2020} studied rapidly spinning compact stars with EoSs featuring a first-order phase transition between strongly coupled nuclear matter and deconfined quark matter and also compatible with the interpretation that the secondary component in GW190814 is a NS. Cao \textit{et al.}~\cite{Cao-2020} provided circumstantial evidence suggesting the recently reported GW190814's secondary component could be an up-down quark star. Dexheimer \textit{et al.}~\cite{Dexheimer-2020} showed that \emph{state-of-the-art} relativistic mean field models can generate massive stars reaching $\geq 2.05 \  M_{\odot}$, while being in good agreement with gravitational-wave events and x-ray pulsar observations when quark vector interactions and higher-order self-vector interactions are introduced. Roupas \textit{et al.}~\cite{Roupas-2020} showed not only that a color-flavor locked quark star with this mass is viable, but also they calculated the range of the model parameters, namely the color superconducting gap  $\Delta$  and the bag constant  $B$ , that satisfies the strict LIGO constraints on the EoS. 

In the study of Moffat~\cite{Moffat-2020},  the modified gravity (MOG) theory is applied to the gravitational wave binary merger GW190814. He  demonstrated that the modified Tolman-Oppenheimer-Volkoff (TOV) equation for a NS can produce a mass  $M=2.5-2.7 \ M_{\odot}$ allowing for the binary secondary component to be identified as a heavy NS in the hypothesized mass gap $M=2.5-5 \ M_{\odot}$. Astashenok \textit{et al.}~\cite{Oikonomou-2020} showed that a NS with this observed mass can be consistently explained with the mass-radius relation obtained by extended theories of gravity. Nunes \textit{et al.}~\cite{Nunes-2020} found that from an appropriate and reasonable combination of modified gravity, rotation effects, and realistic soft EoSs, it is possible to achieve high masses and explain GW190814 secondary component.

Another issue  worth mentioning is the transition of hadron matter to unconfined quark matter at a sufficiently high  density (a few times the saturation density, see previous discussion). Recently, Annala \textit{et al.}~\cite{Annala-2020} claimed that the recent observation of gravitational waves from NSs merger could shed light on the possibility of hadrons to quark phase transition. Moreover, the emergence of strange hadrons (hyperons, etc.) around twice the nuclear saturation density, leads to an appreciable softness of the EoS, and consequently, in most cases the observed values of high mass NSs never reached. As NSs provide a rich testing ground for microscopic theories of dense nuclear matter, combining this study with the experimental data from ultra-relativistic heavy ion collisions may help significantly to improve our knowledge on phase transition theory in hadronic matter. In particular, the so-called Hyperon Puzzle may be addressed (or even more solved)  in NS studies. To be more specific, following the discussion of Ref.~\cite{Bombaci-2017}, the hyperon puzzle is related to the  difficulty to reconcile the measured masses of NSs with the presence of hyperons in their interiors. The presence of hyperons in the interior of NSs is due to the fermionic nature of nucleons. The chemical potential of neutron and proton increasing rapidly, as a function of the density. When the chemical potential of neutrons becomes sufficiently large, the most energetic neutrons can decay via the weak interaction into $\Lambda$ hyperons and form and new Fermi sea for this hadronic species. Other hyperons can be formed with similar weak processes~\cite{Bombaci-2017}. However, the inclusion of hyperons in NS matter was found that leads to an appreciable decrease of maximum NS mass, incompatible with the recent observations. It is stated that this is a common feature of various hyperon star structure calculations (see Ref.~\cite{Bombaci-2017} and reference therein). Thus, although  the presence of hyperons in NSs seems unavailable,  their presence leads to low values of NS mass,  far from observation. This problem is briefly summarized as hyperon puzzle. Of course, there are other studies where the authors  stated that hyperon consideration on  the EoS is not in contradiction with the predictions of very high NS masses (see Ref.~\cite{Sedrakian-2020}). In the the present study, we do not consider the case of additional degrees of freedom (hyperons, quarks, etc.) in the interior of NSs. In any case, additional theoretical calculation in combination with  specific observation  may lead to the solution of the hyperon puzzle and the reveal of  the  existence of  free quark matter in the interior of NSs.

In the present work we concentrate our study on the case where the EoS of NS matter is pure hadronic and the hydrostatic equilibrium is described by the general relativistic equations (TOV equations)~\cite{Shapiro-1983,Glendenning-2000,Haensel-2007,Weinberg-1972,Zeldovich-1971}. Moreover, we focus on two possible cases, that the second partner to be a slow rotating NS and, in the extreme case, to be a very rapidly rotating one (even close to the Kepler limit). Firstly, we consider that the dense nuclear matter properties are described by the MDI-APR~\cite{Koliogiannis-2020} (MDI: momentum dependent interaction, APR: Akmal, Pandharipande and Ravenhall) nuclear model. This model has recently been applied successfully in similar  studies and, according to our opinion, is a robust guide for NS studies. However, since the behavior of dense nuclear matter at high densities remains uncertain, the parametrization of the EoS via the speed of sound is almost inevitable, at least in the framework of hadronic EoSs. To be more specific, we construct large number of EoSs where for low densities (concerning the NS crust), we employ well established results, but for the EoS concerning the core, we apply a twofold consideration. For the outer part we employ the EoS predicted by the MDI-APR model and for the inner one, a parametrization based on the speed of sound  upper limits is applied. In this study, the transition density and the upper limit of the speed of sound are the two free parameters.

In the first part of our study we concentrate on the effect of the speed of sound and transition density on the bulk NS (non-rotating and rapidly rotating) properties including the maximum mass, the Kepler frequency, the kerr parameter, and the maximum central density. We explore under which circumstances the prediction of the mass range $2.5-2.67~M_{\odot}$ of the second partner is possible. More  importantly, we provide the constraints which are inferred by the above consideration.

In the second part of the paper we systematically study the tidal deformability of NSs by employing the large set of EoSs. We mainly focus in the case of high mass candidates existing as an individual star or as a partner in binary NSs system. Until now, there are no observations of an individual, or as a partner of binary system, very massive NS (close to $2.5~M_{\odot}$). However, we consider that it is worth to examine this possibility, by focusing on the predictions of   the tidal deformability and the radius, quantities that are very sensitive on the NS EoS. These quantities are excellent counters on dense matter properties. In the present work, we provide predictions about both the individual and averaged tidal deformability of a hypothetical binary NS system where the most massive partner has a mass in the region $2.5-2.67 \ M_{\odot}$. 

The article is organized as follows: In Section~\ref{sec:mdi+APR}, we present the MDI-APR nuclear model along with quantities at the rotating configuration while in Section~\ref{sec:sos}, we present the speed of of sound parametrization of the EoS. In Section~\ref{sec:tidal}  we provide the basic formalism for the tidal deformability. The results and the discussion are provided in Section~\ref{sec:results} while Section~\ref{sec:remarks} includes the concluding remarks of the present study. Finally, Section~\ref{sec:methods} contains information about the rotating configuration code.

\section{The MDI-APR model and the rapidly rotating neutron star} \label{sec:mdi+APR}
The structure of the EoS and the properties of NSs are studied under the MDI model. In this model, the energy per particle is available through the form~\cite{Prakash-1997,Moustakidis-15}

\begin{eqnarray}
	\label{e-T0}
	E(n,I)&=&\frac{3}{10}E_F^0u^{2/3}\left[(1+I)^{5/3}+(1-I)^{5/3}\right]+
	\frac{1}{3}A\left[\frac{3}{2}-X_{0}I^2\right]u
	+
	\frac{\frac{2}{3}B\left[\frac{3}{2}-X_{3}I^2\right]u^{\sigma}}
	{1+\frac{2}{3}B'\left[\frac{3}{2}-X_{3}I^2\right]u^{\sigma-1}}
	\nonumber \\
	&+&\frac{3}{2}\sum_{i=1,2}\left[C_i+\frac{C_i-8Z_i}{5}I\right]\left(\frac{\Lambda_i}{k_F^0}\right)^3
	\left(\frac{\left((1+I)u\right)^{1/3}}{\frac{\Lambda_i}{k_F^0}}-
	\tan^{-1} \frac{\left((1+
		I)u\right)^{1/3}}{\frac{\Lambda_i}{k_F^0}}\right)\nonumber \\
	&+&
	\frac{3}{2}\sum_{i=1,2}\left[C_i-\frac{C_i-8Z_i}{5}I\right]\left(\frac{\Lambda_i}{k_F^0}\right)^3
	\left(\frac{\left((1-I)u\right)^{1/3}}{\frac{\Lambda_i}{k_F^0}}-
	\tan^{-1}
	\frac{\left((1-I)u\right)^{1/3}}{\frac{\Lambda_i}{k_F^0}}\right),
	\label{eq:mdi_model}
\end{eqnarray}
where $u=n/n_s$, with $n_s$ denoting the saturation density $(n_{s}=0.16~{\rm fm^{-3}})$, $I=(n_{n}-n_{p})/n$ is the asymmetry parameter, $X_{0} = x_{0} + 1/2$, and $X_{3} = x_{3} + 1/2$. The parameters $A$, $B$, $\sigma$, $C_1$, $C_2$, and $B'$ appear in the description of symmetric nuclear matter (SNM) and are determined so that the relation $E(n_{s},0) = -16~{\rm MeV}$ holds. $\Lambda_{1}$ and $\Lambda_{2}$ are finite range parameters equal to $1.5k_{F}^{0}$ and $3k_{F}^{0}$, respectively, with $k_{F}^{0}$ being the Fermi momentum at the saturation density. The rest of the parameters, $x_0$, $x_3$, $Z_1$, and $Z_2$ appear in the description of asymmetric nuclear matter and, with a suitable parametrization, are used in order to obtain different forms for the
density dependence of symmetry energy as well as the value
of slope parameter L and the value of symmetry energy $\rm S_{2}$ at
the saturation density, defined as~\cite{Koliogiannis-2020}
\begin{equation}
	L=3n_{s}\frac{dS_{2}(n)}{dn}\bigg\vert_{n_{s}} \quad \text{and} \quad S_{2}(n)=\frac{1}{2\!}\frac{\partial^{2} E(n,I)}{\partial I^{2}}\bigg\vert_{I=0},
\end{equation}
and as a consequence different parametrizations of the EoS stiffness. In fact, for a specific value of L, the density dependence of symmetry energy is adjusted so that the energy of pure neutron matter is comparable with
those of the existing \emph{state-of-the-art} calculations.

The MDI model, that combines both density and momentum dependent interaction among the nucleons, is suitable for studying NS matter at zero  (present study) as well as at finite temperature. In particular, although it was introduced by Gale \textit{et al.}~\cite{Gale-1987,Gale-1990} to examine the influence of momentum dependent interactions on the momentum flow of heavy-ion collisions, the model has been modified, elaborated, and applied also in the study of the properties of nuclear matter at NSs. The advantages of the MDI model are: (a) reproduces with high accuracy the properties of SNM at the saturation density, including isovector quantities, (b) reproduces the microscopic properties of the Chiral model for pure neutron matter and the results of \emph{state-of-the-art} calculations of Akmal \textit{et al.}~\cite{Akmal-1998}, (c) predicts maximum NS mass higher than the observed ones~\cite{Antoniadis-2013,Cromartie-2016,Linares-2018}, and (d) maintains the causal behavior of the EoS even at densities higher than the ones at the maximum mass configuration.

In this work we apply the EoS produced in Ref.~\cite{Koliogiannis-2020}, where for the construction of the EoS, the MDI model and data from Akmal \textit{et al.}~\cite{Akmal-1998} had been used (for more details see Ref.~\cite{Koliogiannis-2020}). This EoS, not only has the mentioned advantages, but also reproduces the mass of the second component of GW190814 event.

In addition, as a possible scenario is the rotation, we apply rotating configuration in the EoS. In fact, we are interested about the Kepler frequency and the maximum mass of the NS at this configuration. This frequency is considered as the one where the star would shed matter at its equator and consequently is the maximum one (mass-shedding limit). An interesting quantity, which connects the gravitational mass with the angular momentum of the star, is the kerr parameter defined as

\begin{equation}
	\mathcal{K} = \frac{cJ}{GM^2},
	\label{eq:kerr_parameter}
\end{equation}
where M and J are the gravitational mass and angular momentum, respectively. For the construction of the rotating equilibrium model we used the RNS code~\cite{Stergioulas-1996}.

Furthermore, in Ref.~\cite{Breu-2016} had been found an analytical relation which connects the kerr parameter with the gravitational mass of the non-rotating case (TOV), expressed as

\begin{equation}
	M_{\rm rot}=M_{\rm TOV}\left( 1+a_1\left( \frac{\cal K}{{\cal K_{\rm max}}} \right)^2+a_2\left( \frac{\cal K}{{\cal K_{\rm max}}}  \right)^4\right),
	\label{eq:mass_rot_kerr}
\end{equation}
with $a_{1} = 0.132$, $a_{2} = 0.071$, and $\mathcal{K}_{\rm max}$ being the kerr parameter at mass-shedding limit, which is used by the authors to imply constraints on the possible kerr parameter of the second component, as well as the upper limit of a NS mass~\cite{Most-2020}.

\section{Speed of sound formalism and stiffness of equation of state} \label{sec:sos}
Another scenario that is followed to reproduce the mass of the second component is the possible stiffness of the EoS. This is achievable by studying the upper and lower limit on the speed of sound, as well as the possible transition density. In this consideration, we have parametrized the EoS, according to Refs.~\cite{Margaritis-2020,Rhoades-1974,Kalogera-1996,Koranda-1977,Chamel-2013a,Alsing-2018,Podkowka-2018,Xia-2019}, as
\begin{eqnarray}
	P({\cal E})&=&\left\{
	\begin{array}{ll}
		P_{\rm crust}({\cal E}), \quad  {\cal E} \leq {\cal E}_{\rm c-edge}  & \\
		\\
		P_{\rm NM}({\cal E}), \quad  {\cal E}_{\rm c-edge} \leq {\cal E} \leq {\cal E}_{\rm tr}  & \\
		\\
		\left(\frac{v_{\rm s}}{c}  \right)^2\left({\cal E}-{\cal E}_{\rm tr}  \right)+
		P_{\rm NM}({\cal E}_{\rm tr}), \quad  {\cal E}_{\rm tr} \leq {\cal E} , & \
	\end{array}
	\right.
	\label{eq:eos_sos}
\end{eqnarray}
where $P$ and $\mathcal{E}$ denote the pressure and energy density, respectively, and $\mathcal{E}_{\rm tr}$ is the transition energy density. For the construction of the EoSs, we adopted the following: (a) in region ${\cal E} \leq {\cal E}_{\rm c-edge}$, we used the equation of Feynman {\it et al.}~\cite{Feynman-1949} and also of Baym {\it et al.}~\cite{Baym-1971} for the crust and low densities of NS, (b) in the intermediate region, ${\cal E}_{\rm c-edge} \leq {\cal E} \leq {\cal E}_{\rm tr}$, we employed a specific EoS based on the MDI model and data from Akmal {\it et al.}~\cite{Akmal-1998}, and (c) for ${\cal E}_{\rm tr}\geq \mathcal{E}$ region, the EoS is maximally stiff with the speed of sound, defined as  $v_s=c\sqrt{\left(\partial P / \partial {\cal E}\right)}_{\rm S}$ (where $S$ is the entropy) fixed in the present work in the range $[c/\sqrt{3},c]$. The lowest allowed
value of the speed of sound, that is $(v_{s}/c)^{2}=1/3$, is introduced in order to be
consistent with the possibility of a phase transition in quark matter. In this
case, all the theoretical predictions lead to this value as an upper bound of
the speed of sound. The implementation of speed of sound values between the limited ones will lead to results well constrained by the two mentioned limits. Although the energy densities below the ${\cal E}_{\rm c-edge}$ have negligible effects on the maximum mass configuration, we used them in calculations for the accurate estimation of the tidal deformability. 
The cases which took effect in this study can be divided into two categories based on the fiducial baryon transition density, $n_{\rm tr} = p n_{\rm s}$, and the speed of sound as: (a) the ones where $p$ takes the values $[1.5,2,3,4,5]$, while the speed of sound is parametrized in the two limiting cases, $(v_{s}/c)^{2}=1/3$ and $(v_{s}/c)^{2}=1$ and (b) the ones where $p$ takes the values $[1.5,2]$, while the speed of sound is parametrized in the range $(v_{s}/c)^{2}=[1/3,1]$. The predicted EoSs are functional of $n_{\rm tr}$ and $(v_s/c)$ and implemented to study the possibly existence of a NS with $\sim 2.6~M_{\odot}$, either non-rotating or a rotating one.

In approach followed in Eq.~\eqref{eq:eos_sos}, while the continuity on the EoS is well ensured, the continuity in the speed of sound at the transition density, due to its artificial character, is not.  Therefore, in order to ensure the continuity and a smooth phase transition, we employ a method presented in Ref.~\cite{Tews-2018}. We proceeded with the matching of the EoSs on the transition density  by considering that, above this value, the speed of sound is parametrized as follows (for more details see Ref.~\cite{Tews-2018})
\begin{equation}
	\frac{v_{\rm s}}{c}=\left(a-c_1\exp\left[-\frac{(n-c_2)^2}{w^2} \right]\right)^{1/2}, \quad a\in [1/3,1]
	\label{speed-matc-1}
\end{equation}
where the parameters $c_1$ and $c_2$ are fit to the speed of sound and its derivative at $n_{\rm tr}$, and also to the demands $v_{\rm s}(n_{\rm tr})=[c/\sqrt{3},c]$~\cite{Margaritis-2020} according to the value of $\alpha$. The remaining parameter $w$ controls the width of the curve, where in our case is equal to $10^{-3}~{\rm fm^{-3}}$ in order to preserve the NS properties. Using Eq.~(\ref{speed-matc-1}), the EoS for $n \geq n_{{\rm tr}}$ can be constructed with the help of the following recipe~\cite{Tews-2018}
\begin{equation}
	{\cal E}_{i+1} = {\cal E}_i+\Delta {\cal E}, \quad P_{i+1} = P_i+\left(\frac{v_s}{c}(n_i)\right)^2\Delta {\cal E},
	\label{eq:5}
\end{equation}
\begin{equation}
	\Delta {\cal E} = \Delta n\left(\frac{{\cal E}_i+P_i}{n_i} \right),
	\label{eq:6}
\end{equation} 
\begin{equation}
	\Delta n = n_{i+1}-n_i.
	\label{eq:7}
\end{equation}
The treatment with both discontinuity and continuity in the speed of sound is presented in Table V of Ref.~\cite{Margaritis-2020}. The outline was that the two approaches converge and consequently the effects of the discontinuity are negligible.

\begin{figure}[h!]
	\centering
	\includegraphics[width=1.0\textwidth]{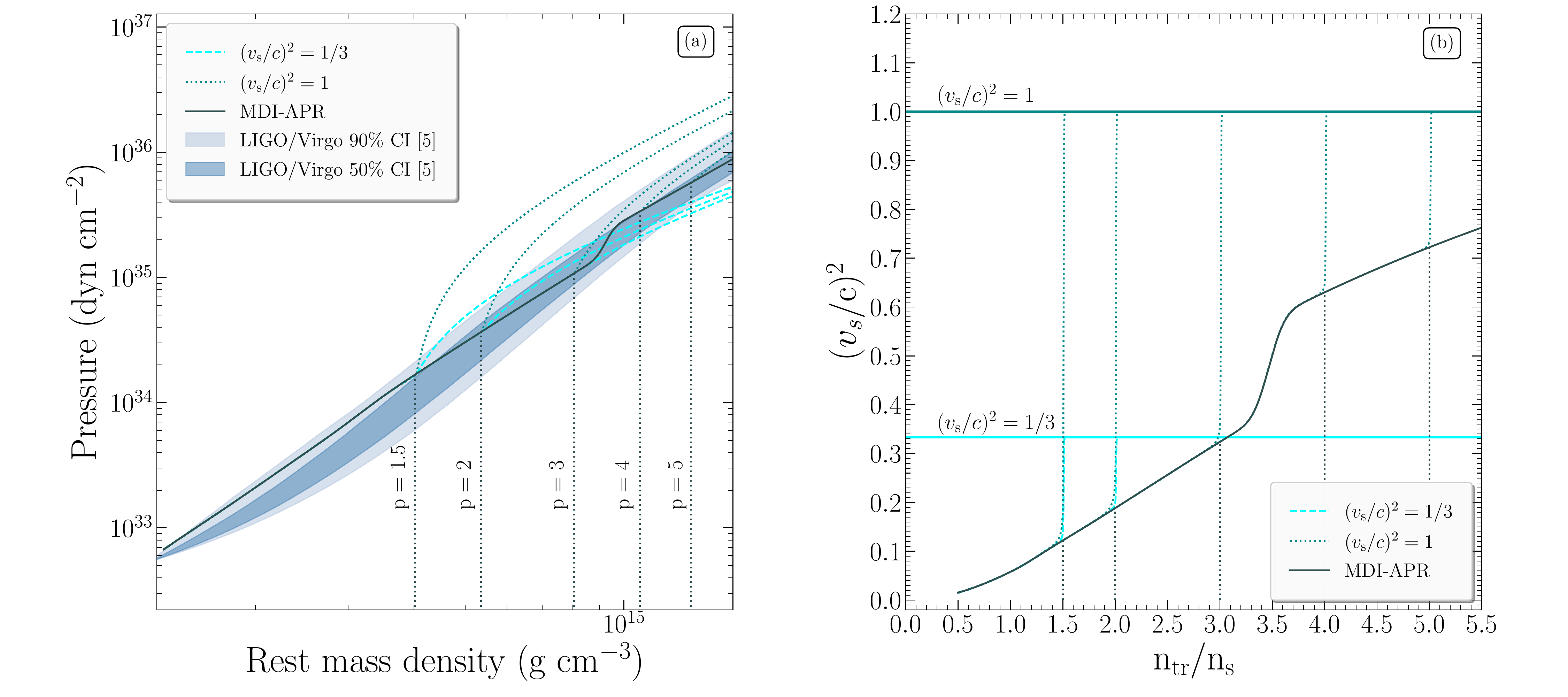}
	\includegraphics[width=1.0\textwidth]{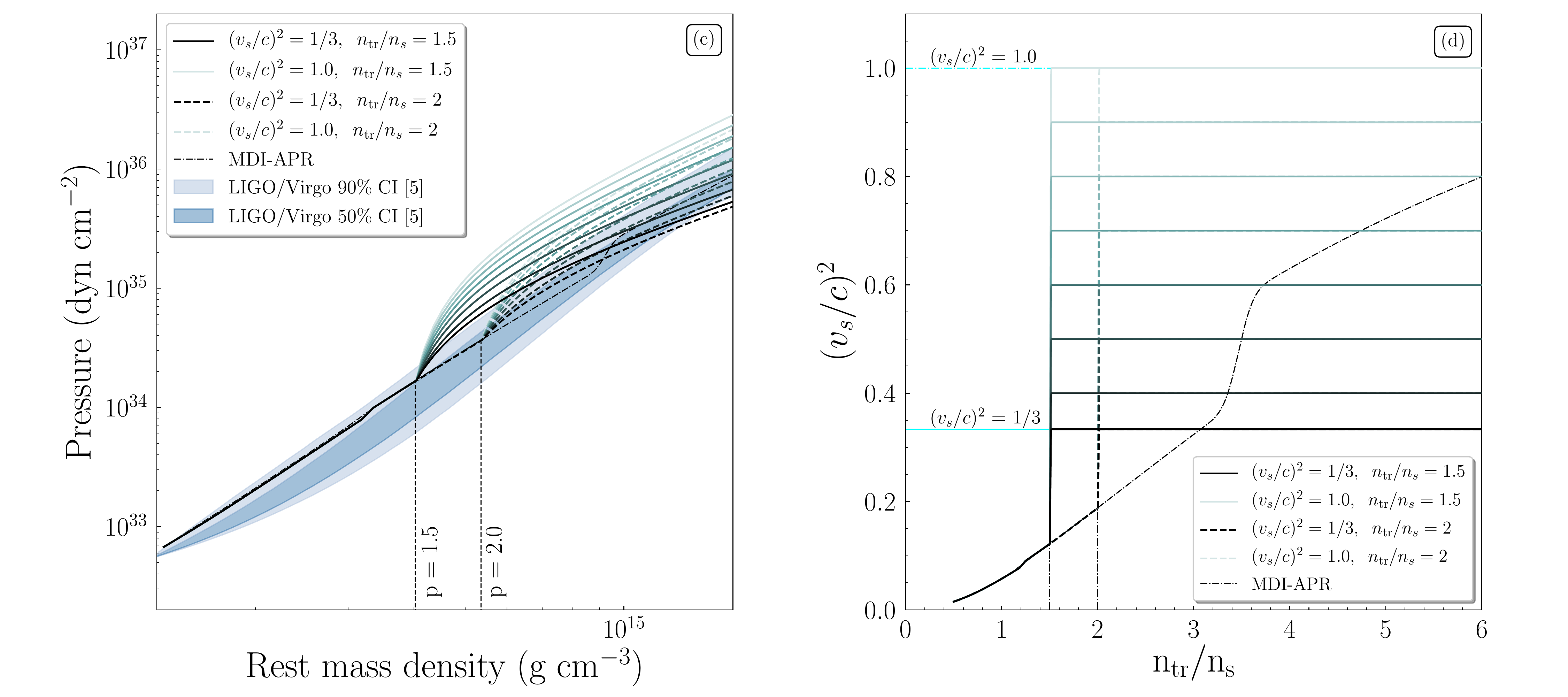}
\caption{(a) Pressure as a function of rest mass density and (b) square speed of sound in units of speed of light as a function of the transition density, where $p$ takes the values $[1.5,2,3,4,5]$, while the speed of sound is parametrized in the two limiting cases, $(v_{s}/c)^{2}=1/3$ and $(v_{s}/c)^{2}=1$. (c) Pressure as a function of rest mass density and (d) square speed of sound in units of speed of light as a function of the transition density, where $p$ takes the values $[1.5,2]$, while the speed of sound is parametrized in the range $(v_{s}/c)^{2}=[1/3,1]$ (As the speed of sound is getting higher values, the curves' color lightens). The vertical lines display the transition cases, while the shaded regions show the credibility interval extracted from Ref.~\cite{Abbott-2018}.}
	\label{fig:pressure_density_sos}
\end{figure}

In Fig.~\ref{fig:pressure_density_sos} we present the pressure as a function of the rest mass density ($\rho_{\rm rest} = n_{b}m_{n}$) and the square speed of sound in units of speed of light as a function of the transition density for the EoSs constructed in cases (a) and (b). In addition, we display the credibility intervals proposed by Ref.~\cite{Abbott-2018} from LIGO/Virgo collaboration for the GW170817 event. It is clear from these figures that the pure MDI-APR EoS is well-defined in the proposed limits of LIGO/Virgo collaboration and also fulfills the speed of light limit at high densities. 

\section{Tidal deformability} \label{sec:tidal}
It has mentioned that the gravitational waves emitted from the final stages of an inspiraling binary NS system are one of the most important sources for the terrestrial gravitational waves detectors~\cite{Postnikov-2010,Baiotti-2019,Flanagan-08,Hinderer-08,Damour-09,Hinderer-10,Fattoyev-13,Lackey-015,Takatsy-2020}. In such case, properties like the mass of the component stars can be measured. As  Flanagan and Hinderer \cite{Flanagan-08} articulated, the tidal effects can be measurable during this final stage of the inspiral.

The response of a NS to the presence of the tidal field, is described by a dimensionless tidal parameter, the tidal Love number $k_2$. This parameter depends on the NS structure; hence its mass and  EoS. The tidal Love number $k_2$ is the coefficient of proportionality between the induced quadrupole moment $Q_{ij}$ and the applied tidal field $E_{ij}$~\cite{Flanagan-08,Thorne-1998}, given below

\begin{equation}
Q_{ij}=-\frac{2}{3}k_2\frac{R^5}{G}E_{ij}\equiv- \lambda E_{ij},
\label{Love-1}
\end{equation}
where $R$ is the NS radius and $\lambda=2R^5k_2/3G$ is the other tidal parameter that we use in our study, the so-called tidal deformability. The tidal Love number $k_2$ is given by \cite{Flanagan-08,Hinderer-08}
\begin{eqnarray}
k_2&=&\frac{8\beta^5}{5}\left(1-2\beta\right)^2\left[2-y_R+(y_R-1)2\beta \right] \times
\left[\frac{}{} 2\beta \left(6  -3y_R+3\beta (5y_R-8)\right) \right. \nonumber \\
&+& 4\beta^3 \left.  \left(13-11y_R+\beta(3y_R-2)+2\beta^2(1+y_R)\right)\frac{}{} \right.\nonumber \\
&+& \left. 3\left(1-2\beta \right)^2\left[2-y_R+2\beta(y_R-1)\right] {\rm ln}\left(1-2\beta\right)\right]^{-1},
\label{k2-def}
\end{eqnarray}
where $\beta=GM/Rc^2$ is the compactness parameter of a NS. The quantity $y_R$ is determined by solving the following differential equation
\begin{equation}
r\frac{dy(r)}{dr}+y^2(r)+y(r)F(r)+r^2Q(r)=0, 
\label{D-y-1}
\end{equation}
with the initial condition $ y(0)=2$~\cite{Hinderer-10}. $F(r)$ and $Q(r)$ are functionals of ${\cal E}(r)$, $P(r)$ and $M(r)$  defined as~\cite{Postnikov-2010,Hinderer-10}
\begin{equation}
F(r)=\left[ 1- \frac{4\pi r^2 G}{c^4}\left({\cal E} (r)-P(r) \right)\right]\left(1-\frac{2M(r)G}{rc^2}  \right)^{-1},
\label{Fr-1}
\end{equation}
and
\begin{eqnarray}
r^2Q(r)&=&\frac{4\pi r^2 G}{c^4} \left[5{\cal E} (r)+9P(r)+\frac{{\cal E} (r)+P(r)}{\partial P(r)/\partial{\cal E} (r)}\right]
\times\left(1-\frac{2M(r)G}{rc^2}  \right)^{-1}- 6\left(1-\frac{2M(r)G}{rc^2}  \right)^{-1} \nonumber \\
&-&\frac{4M^2(r)G^2}{r^2c^4}\left(1+\frac{4\pi r^3 P(r)}{M(r)c^2}   \right)^2\left(1-\frac{2M(r)G}{rc^2}  \right)^{-2}.
\label{Qr-1}
\end{eqnarray}

The numerical solution requires that the Eq.~(\ref{D-y-1})  must be integrated self consistently  with the TOV equations using the boundary conditions $y(0)=2$, $P(0)=P_c$ and $M(0)=0$~\cite{Postnikov-2010,Hinderer-08}. From the solution of TOV equations the mass $M$ and radius $R$ of the NS can be extracted, while the corresponding solution of the differential Eq.~(\ref{D-y-1}) provides the value of $y_R=y(R)$. This parameter along with the quantity $\beta$ are the  basic ingredients  of the tidal Love number $k_2$.

One parameter that is well constrained by the gravitational waves detectors is the chirp mass {\it $\mathcal{M}_c$}, which is a combination of the component masses of a binary NS system~\cite{Abbott-1,Abbott-3}
\begin{equation}
\mathcal{M}_c=\frac{(m_1m_2)^{3/5}}{(m_1+m_2)^{1/5}}=m_1\frac{q^{3/5}}{(1+q)^{1/5}},
\label{chirpmass}
\end{equation}
where $m_1$ is the mass of the heavier component star and $m_2$ is the lighter's one. Hence, the binary mass ratio $q=m_2/m_1$ is within the range $0\leq q\leq1$.

In addition, another binary parameter that can be measured from the analysis of the gravitational wave signal is the effective tidal deformability~\cite{Abbott-1,Abbott-3}
\begin{equation}
\tilde{\Lambda}=\frac{16}{13}\frac{(12q+1)\Lambda_1+(12+q)q^4\Lambda_2}{(1+q)^5},
\label{L-tild-1}
\end{equation}
where the key quantity $q$ characterizes the mass asymmetry, and $\Lambda_i$ is the dimensionless deformability defined as~\cite{Abbott-1,Abbott-3}
\begin{equation}
\Lambda_i=\frac{2}{3}k_2\left(\frac{R_i c^2}{M_i G}  \right)^5\equiv\frac{2}{3}k_2 \beta_i^{-5}  , \quad i=1,2.
\label{Lamb-1}
\end{equation}
By combining Eq.~(\ref{Lamb-1}) with  Eq.~(\ref{k2-def}), one can find that  $\Lambda_i$  depends both on star's compactness  and the value of $y(R)$. 

We notice that $\Lambda_i$  depends directly on the stiffness of the EoS through the compactness $\beta$ and  indirectly through the speed of sound which appears in Eq.~(\ref{Qr-1}). The dependence of $\Lambda_i$ on the behavior of $y(R)$ can lead to useful estimations or constraints on the tidal deformability itself. To be more specific, the applied EoS affects also the behavior of $\Lambda$ regarding the NS's  mass $M$ and radius $R$ . In our study we use the secondary very massive component of GW190814 system (see Ref.~\cite{Abbott-4}) to examine the tidal deformability and the behavior of dense nuclear matter in the extreme scenario of such a massive NS.  

\section{Results and Discussion} \label{sec:results}
The merger of a very massive black hole ($\sim 23~M_{\odot}$) with a $\sim 2.6~M_{\odot}$ compact object has recently announced by the LIGO/Virgo collaboration, as the GW190814 event. The scenarios that follow the second merger component are that of (a) the lightest black hole, (b) the most compact NS, (c) a rapidly rotating NS, and (d) an exotic compact object. In the present work we study only the second and third case scenarios, that is either a compact non-rotating NS or a rapidly rotating one.

\subsection{Slow/rapid rotation: Implications to neutron star properties}

\begin{figure}[h!]
	\centering
	\includegraphics[width=0.5\textwidth]{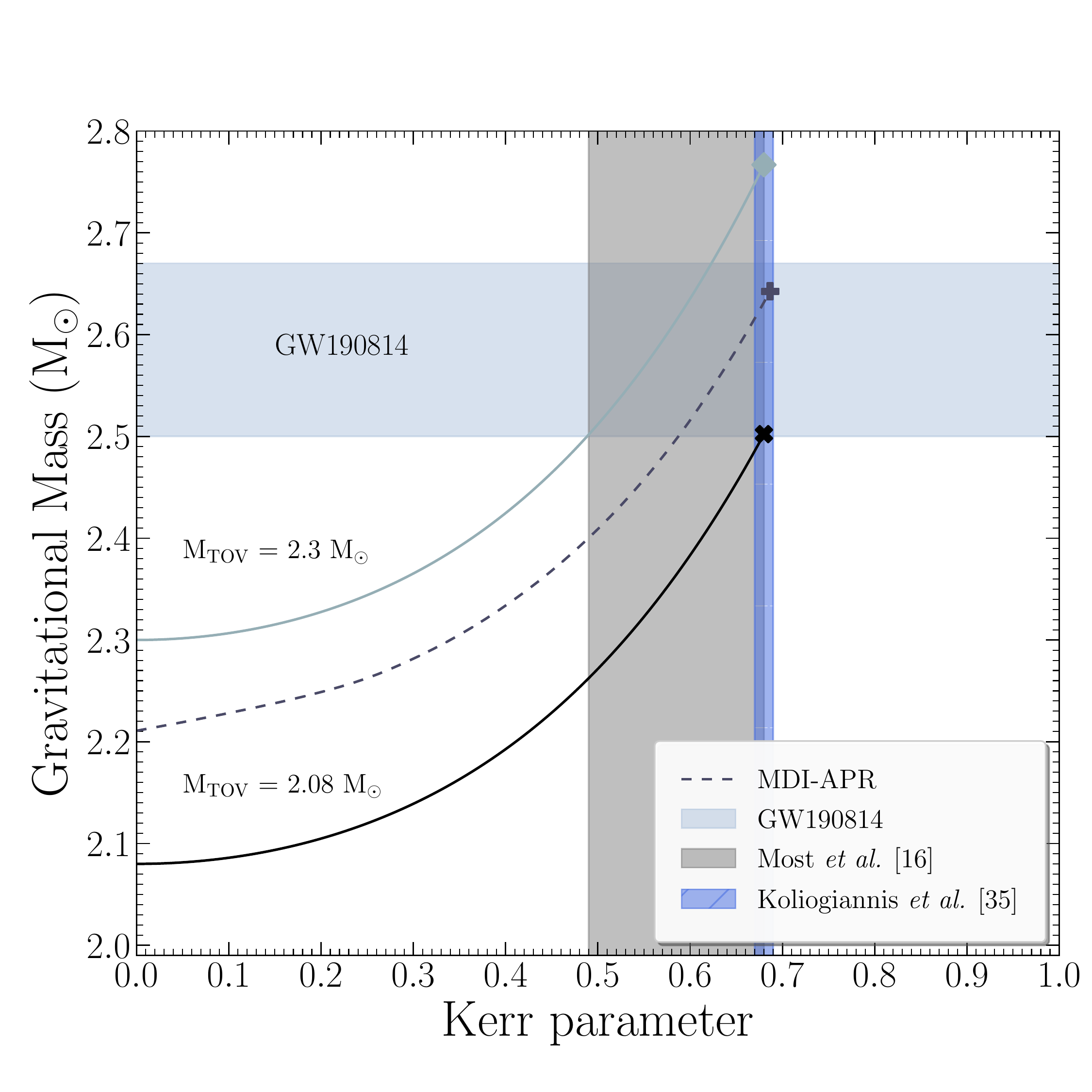}
	\caption{Gravitational mass as a function of kerr parameter for the MDI-APR EoS. The solid lines from bottom to top represent the Eq.~\eqref{eq:mass_rot_kerr} with $M_{\rm TOV} = 2.08~M_{\odot}$ and $M_{\rm TOV} = 2.3~M_{\odot}$. The mass range of the second component of GW190814 is noted with the horizontal shaded region, while with the vertical one (left), the possible region of kerr parameter $\mathcal{K} = [0.49,0.68]$ from Ref.~\cite{Most-2020} is shown. In addition, the region for the kerr parameter $\mathcal{K}_{\rm max} = [0.67,0.69]$ from Ref.~\cite{Koliogiannis-2020}, if the low mass component was rotating at its mass-shedding limit, is presented with the vertical shaded region (right). The markers point the maximum mass configuration at the mass-shedding limit.}
	\label{fig:mass-kerr}
\end{figure}

In Fig.~\ref{fig:mass-kerr} we display the gravitational mass as a function of the kerr parameter for the pure MDI-APR EoS. In addition, we note the universal relation Eq.~\eqref{eq:mass_rot_kerr} for two limiting cases: (a) $M_{\rm TOV} = 2.08~M_{\odot}$ and (b) $M_{\rm TOV} = 2.3~M_{\odot}$~\cite{Most-2020}, and $\mathcal{K}_{\rm max} = 0.68$. The limiting cases are the minimum and maximum possible mass, respectively, for a NS based on the calculations provided in Ref.~\cite{Most-2020}. In accordance, the maximum value of the kerr parameter is also calculated in Ref.~\cite{Most-2020} with respect to the minimum possible mass. With regard to the pure MDI-APR EoS, the relevant dependence is constructed through the RNS code having as input the angular momentum of the star until it reaches its mass-shedding limit. This figure represents the limited area where the compact object should lie. The area is marked by the intersection of the gravitational mass, $M=2.59^{+0.08}_{-0.09}~M_{\odot}$, with the kerr parameter, $\mathcal{K}=[0.49,0.68]$~\cite{Most-2020}. We note that the pure MDI-APR EoS is in the range of the described limits for the gravitational mass and kerr parameter, as well as the ones introduced in Fig.~\ref{fig:pressure_density_sos}, being a suitable hadronic EoS to simulate the compact object of $\sim 2.6~M_{\odot}$. 

In addition, taking into consideration the limiting case that the compact object is rotating at its mass-shedding limit, then constraints on the maximum value of the kerr parameter, the corresponding equatorial radius, and the central energy density are possible. In particular, firstly we employ the relation found in Ref.~\cite{Koliogiannis-2020}

\begin{equation}
	\mathcal{K}_{\rm max} = 0.488 + 0.074 \left(\frac{M_{\rm max}}{M_{\odot}}\right),
\end{equation}
for the observable gravitational mass. For the mass of the second component, the kerr parameter lies in the range $\mathcal{K}_{\rm max} = [0.67,0.69]$, which is also noted in Fig.~\ref{fig:mass-kerr}. Secondly, using the derived relation from the recent Ref.~\cite{Koliogiannis-2-2020}, which connects the maximum value of kerr parameter with the one of compactness parameter, as

\begin{equation}
	\mathcal{K}_{\rm max} = 1.34 \sqrt{\beta_{\rm max}} \quad \text{with} \quad \beta_{\rm max} = \frac{G}{c^{2}}\frac{M_{\rm max}}{R_{\rm max}},
\end{equation} 
it is possible to extract a specific range for the corresponding equatorial radius. In this case, the corresponding equatorial radius lies in the range $R_{\rm max} = [14.77,14.87]~{\rm km}$.

\begin{table}[H]
	\centering
		\caption{Coefficients of Eqs.~\eqref{eq:sos_mass} and~\eqref{eq:sos_kerr} for the two speed of sound bounds. The abbreviation ``N.R." corresponds to the non-rotating configuration and the ``M.R." to the maximally-rotating one.}
		\begin{tabular}{ccccccccc}
		\toprule
		\multirow{2}{*}{\textbf{Speed of sound bounds}} & \multicolumn{2}{c}{$\alpha_{1}$} & \multicolumn{2}{c}{$\alpha_{2}$} & \multicolumn{2}{c}{$\alpha_{3}$} & \multicolumn{2}{c}{$\alpha_{4}$}\\
		& \textbf{N.R.} & \textbf{M.R.} & \textbf{N.R.} & \textbf{M.R.} & \textbf{N.R.} & \textbf{M.R.} & \textbf{N.R.} & \textbf{M.R.}\\
		\midrule
		$c$ & 1.665 & 1.689 & 0.448 & 0.352 & -- & 0.683 & -- & 1.053 \\
	
		$c/\sqrt{3}$ & 1.751 & 2.069 & 0.964 & 0.883 & -- & 0.645 & -- & 1.348 \\
		\bottomrule
		\end{tabular}
\label{tab:table3}
\end{table}

\begin{figure}[t!]
	\centering
	\includegraphics[width=1.0\textwidth]{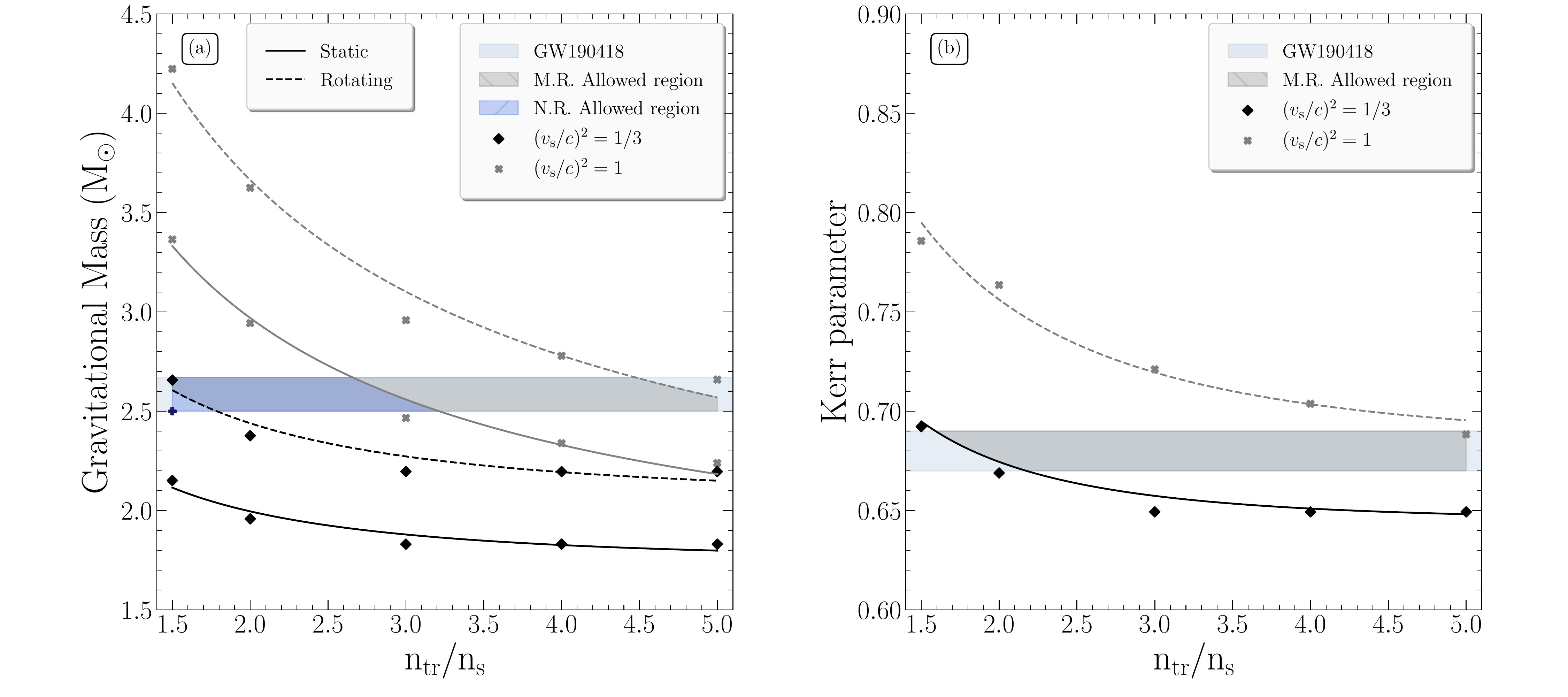}
	\caption{(a) Gravitational mass and (b) kerr parameter as a function of transition density at the maximum mass configuration for the two limiting speed of sound bounds. The data at the maximum mass configuration is presented with diamonds for the $(v_{s}/c)^{2}=1/3$ bound and crosses for the $(v_{s}/c)^{2}=1$ bound. The plus marker denotes the lower bound in the speed of sound, $(v_{s}/c)^{2}=0.45$, assuming that the second component was a non-rotating NS. The mass range of the second component of GW190814 is noted with the horizontal shaded region. (a) The lighter shaded region marks the allowed range for the transition density at the maximally-rotating (M.R.) configuration, while the darker one, marks the allowed region at the non-rotating (N.R.) configuration. (b) The darker shaded region marks the allowed range for the transition density at the maximally-rotating (M.R.) configuration.}
	\label{fig:mass_kerr-density}
\end{figure}

We concentrate now on the microscopic properties of the NS, the speed of sound and the transition density. In Fig.~\ref{fig:mass_kerr-density} we display the gravitational mass and kerr parameter as a function of the transition density for the two limiting cases of the speed of sound based on Ref.~\cite{Margaritis-2020}. From Fig.~\ref{fig:mass_kerr-density}(a), the intersection of GW190814 mass area with the extracted curves provide us two regions of the possible transition density with respect to the applied speed of sound. By employing the formula from Ref.~\cite{Margaritis-2020}

\begin{equation}
	\frac{M_{\rm max}}{M_{\odot}} = \alpha_{1} \coth\left[\alpha_{2}\left(\frac{n_{\rm tr}}{n_s}\right)^{1/2}\right],
	\label{eq:sos_mass}
\end{equation} 
where the coefficients $a_{1}$ and $a_{2}$ are given in Table~\ref{tab:table3}, we were able to restrict the transition density with respect to the speed of sound. More precisely, we took under consideration two possible cases: (a) non-rotating and (b) maximally-rotating NS. In the first case, as Fig.~\ref{fig:mass_kerr-density}(a) shows, the possible transition density region is restricted between the two limiting cases of the speed of sound. However, as the lower limit in the speed of sound is not able to represent the gravitational mass of the low mass component, we found the lower possible speed of sound value that reproduces this mass at the specific value of the transition density. Consequently, the transition density is constrained in the range $n_{\rm tr} = [1.5,3.2]~n_{s}$ and the speed of sound in the range $(v_{s}/c)^{2} = [0.45,1]$.

In the second case, the possible transition density can be constrained both from the gravitational mass and the kerr parameter. From Fig.~\ref{fig:mass_kerr-density}(a) it is clear that the transition density can take all the values in the area under consideration. However, from Fig.~\ref{fig:mass_kerr-density}(b) the transition density is constrained from the lower limit of the speed of sound. According to the derived formula from Ref.~\cite{Margaritis-2020}

\begin{equation}
	\mathcal{K}_{\rm max} = \alpha_{3} \coth\left[\alpha_{4}\left(\frac{n_{\rm tr}}{n_s}\right)^{1/2}\right],
	\label{eq:sos_kerr}
\end{equation}
where the coefficients $a_{3}$ and $a_{4}$ are given in Table~\ref{tab:table3}, the rotating case allows the existence of the low mass component at the whole range of speed of sound values and at transition densities in the range $n_{\rm tr} = [1.6,5]~n_{s}$.

\begin{figure}[h!]
	\centering
	\includegraphics[width=0.5\textwidth]{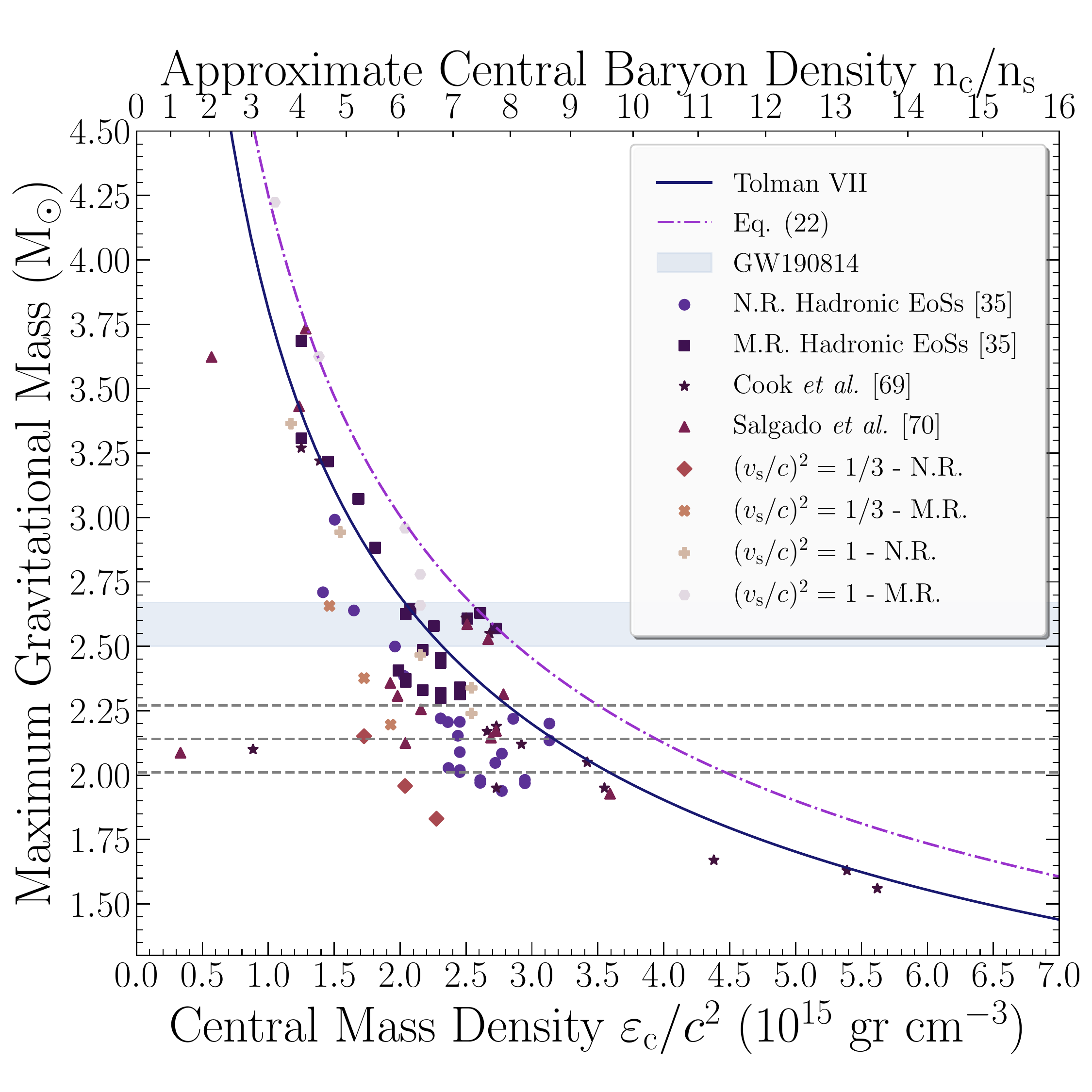}
	\caption{Gravitational mass as a function of the central energy/baryon density at the maximum mass configuration both at non-rotating and maximally-rotating case. Circles correspond to 23 hadronic EoSs~\cite{Koliogiannis-2020} at the non-rotating case (N.R.), squares to the corresponding maximally-rotating (M.R.) one, stars to data of Cook \textit{et al.}~\cite{Cook-1994}, and triangles to data of Salgado \textit{et al.}~\cite{Salgado-1994}. In addition, rhombus and pluses mark the non-rotating configuration at the two limiting values of the sound speed, while crosses and polygons marks the maximally-rotating one. The horizontal dashed lines correspond to the observed NS mass limits ($2.01~M_{\odot}$~\cite{Antoniadis-2013}, $2.14~M_{\odot}$~\cite{Cromartie-2016}, and $2.27~M_{\odot}$~\cite{Linares-2018}). Eq.~\eqref{eq:mass_energy} is noted with the dashed-dotted line, while for comparison the Tolman VII analytical solution~\cite{Koliogiannis-2020} is shown with the solid line. The mass range of the second component of GW190814 is noted with the horizontal shaded region.}
	\label{fig:mass-energy density}
\end{figure}

One more interesting property of NSs is the central energy density, as it is connected with the study of the time evolution of pulsars and the appearance of a possible phase transition. In Ref.~\cite{Koliogiannis-2020} a relation was found describing the upper bound for the density of cold baryonic matter, as

\begin{equation}
	\frac{M}{M_{\odot}} = 4.25 \sqrt{\frac{10^{15}~{\rm gr}~{\rm cm^{-3}}}{\varepsilon_{\rm c}/c^{2}}}.
	\label{eq:mass_energy}
\end{equation}

Fig.~\ref{fig:mass-energy density} presents the maximum gravitational mass as a function both of the central energy density and the central baryon density. In particular, we present the results of 23 hadronic EoSs~\cite{Koliogiannis-2020}, for the non-rotating and maximally-rotating case, Tolman VII analytical solution, Eq.~\eqref{eq:mass_energy}, data from Cook \textit{et al.}~\cite{Cook-1994} and Salgado \textit{et al.}~\cite{Salgado-1994}, as well as the newly added data for the non-rotating and maximally-rotating case, both in $(v_{s}/c)^2=1/3$ and $(v_{s}/c)^2=1$, at the transition densities under consideration. Adopting the Eq.~\eqref{eq:mass_energy} for the range of the gravitational mass of the low mass component in the GW190814 event, the central energy density can be constrained in the range $\varepsilon_{\rm c}/c^{2} = [2.53,2.89]~10^{15}~{\rm gr}~{\rm cm^{-3}}$, meaning that NSs with higher values of central energy density cannot exist. Furthermore, from Fig.~\ref{fig:mass-energy density}, we can also extract the corresponding region for the central baryon density, that is $n_{c} = [7.27,8.09]~n_{s}$. Finally, all the extracted EoSs meet the limit for the central energy/baryon density as they are included in the region described under Eq.~\eqref{eq:mass_energy}.

\subsection{Tidal effects and speed of sound: A very massive neutron star hypothesis}
\subsubsection{Isolated non-rotating neutron star}
\begin{figure}[h!]
	\centering
	\includegraphics[width=0.5\textwidth]{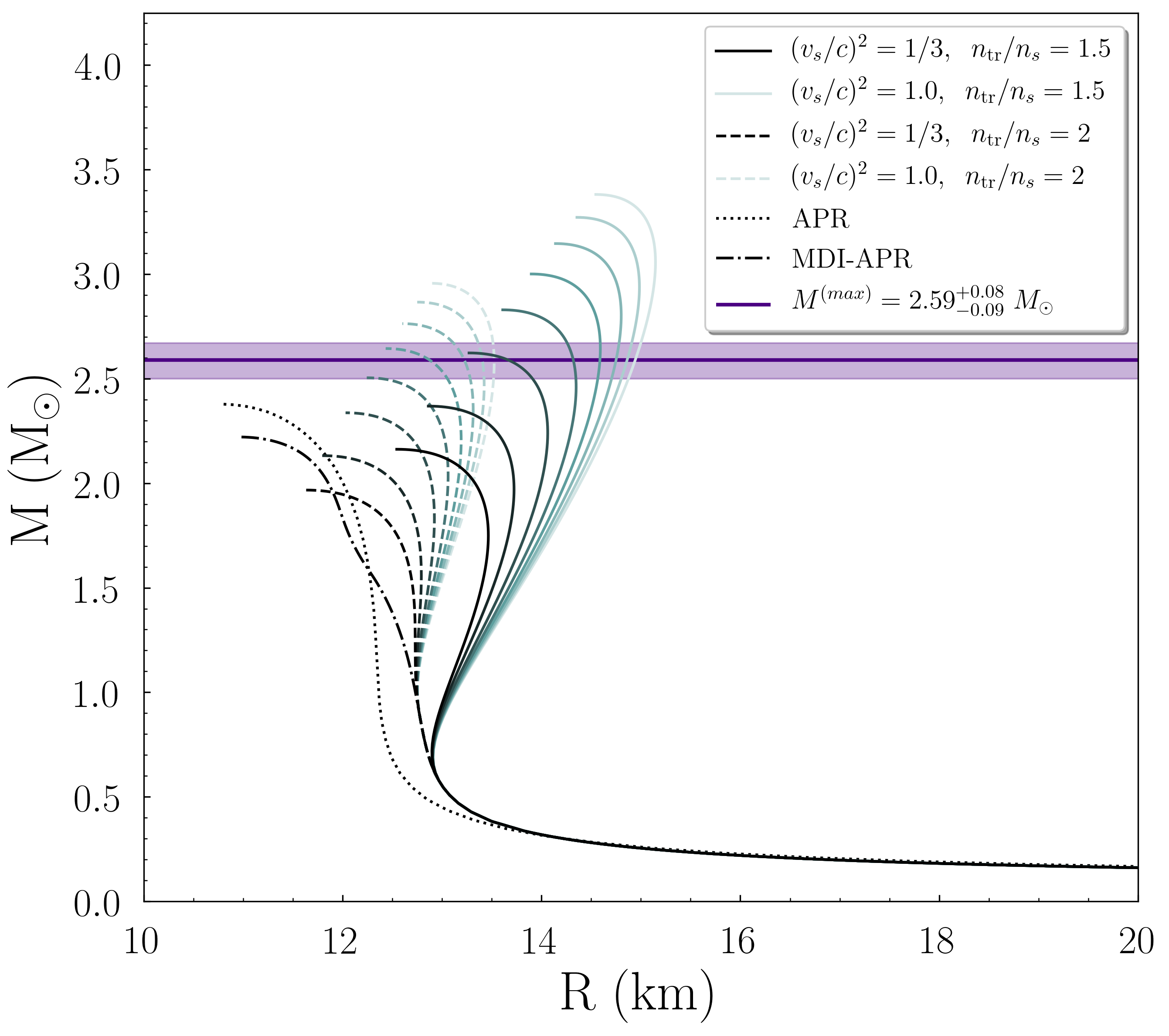}
	\caption{Mass vs. radius for an isolated non-rotating NS, for each transition density $n_{\mathrm{tr}}$ and all speed of sound cases. The higher values of speed of sound correspond to lighter curves' color. The purple horizontal line and region indicate the mass estimation of the massive compact object of Ref.~\cite{Abbott-4}. The dashdot (dotted) curve corresponds to the MDI-APR (APR) EoS.}
	\label{fig:MR}
\end{figure}

Firstly, we concentrated our study of tidal deformability on the isolated non-rotating NS case, by using two transition densities $n_{\mathrm{tr}}=[1.5,2]n_s$ and eight values of speed of sound bounds $(v_s/c)^2=[1/3,0.4,0.5,0.6,0.7,0.8,0.9,1]$. The values of transition density were taken to be close to the constraints of Ref.~\cite{Kanakis-2020}. The numerical solution of TOV equations' system, by using the previous bounds for sound speed, provided the mass-radius diagram presented in Fig.~\ref{fig:MR}.

In Fig.~\ref{fig:MR} one can observe two main branches, related to the transition density; the solid (dashed) curves correspond to the $n_{\mathrm{tr}}=1.5n_s$ ($n_{\mathrm{tr}}=2n_s$) case. In each branch, there are bifurcations in the families of EoSs, in analog to each speed of sound boundary condition. The higher the speed of sound, the lighter the representing color of the curves in the figure. The purple solid horizontal line, with the shaded region, indicates the estimation of the recently observed massive compact object of Ref.~\cite{Abbott-4}. As Fig.~\ref{fig:MR} shows, the branch of EoSs with $n_{\mathrm{tr}}=1.5n_s$ provides stiffer EoSs compare to the $n_{\mathrm{tr}}=2n_s$ branch. The EoSs of the $n_{\mathrm{tr}}=1.5n_s$ case are more likely to provide such a massive non-rotating NS, than the $n_{\mathrm{tr}}=2n_s$ case in which three EoSs of the total sum lie outside of the shaded region. Especially, between the same kind of transition density $n_{\mathrm{tr}}$ the EoSs with higher speed of sound bounds lead to higher values of NS mass and radius, hence a high bound of the speed of sound (even more close to the causality as the transition density is getting higher) is needed for the description of such a massive compact object.

\begin{figure}[h!]
	\centering
	\includegraphics[width=0.5\textwidth]{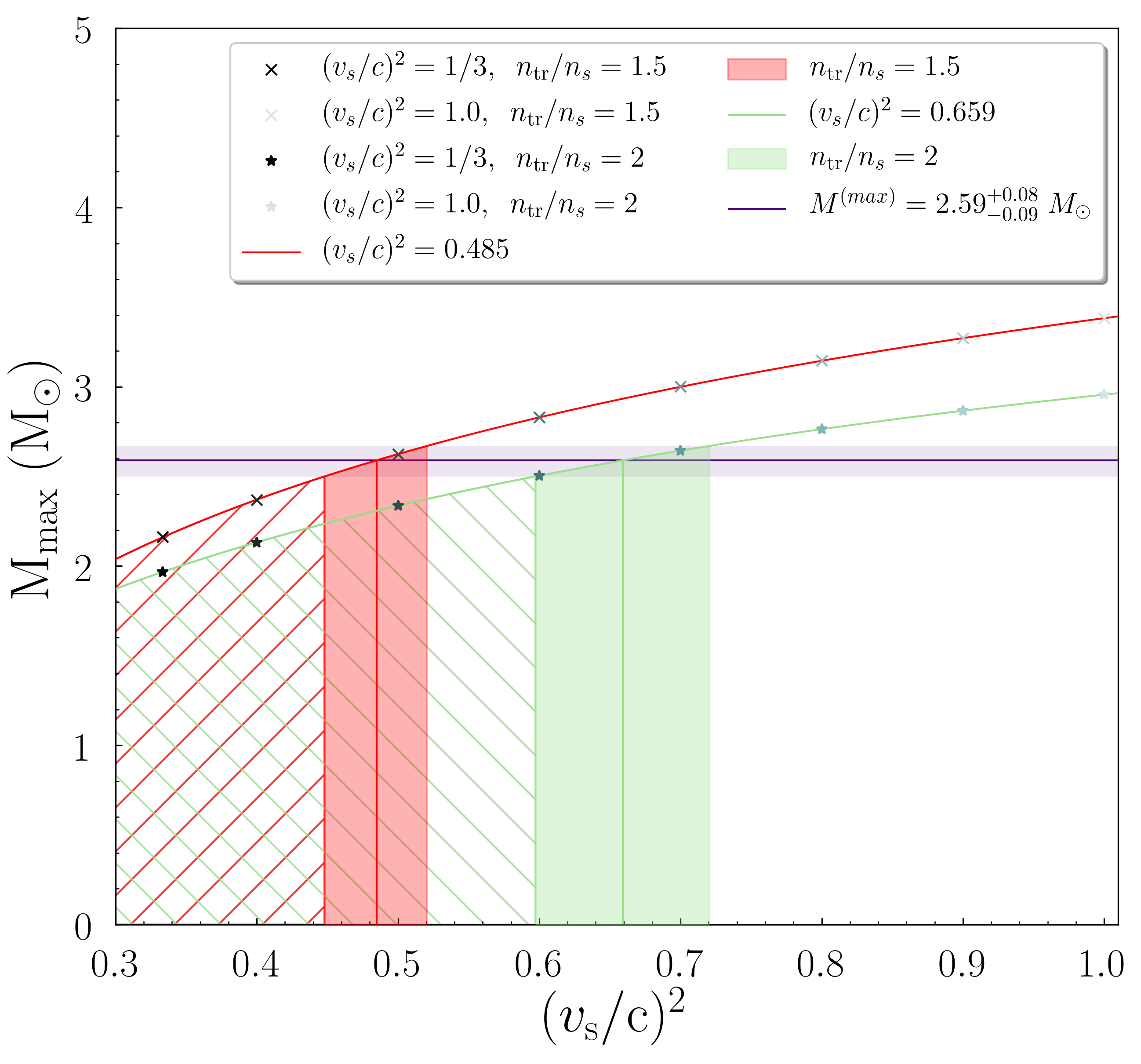}
	\caption{Dependence of a non-rotating NS's maximum mass $M_{max}$ on the speed of sound values $(v_s/c)^2$ for each transition density $n_{\mathrm{tr}}$ (in units of saturation density $n_s$). The red vertical shaded region corresponds to the $n_{\mathrm{tr}}=1.5n_s$ case, while the green one corresponds to the $n_{\mathrm{tr}}=2n_s$ case. The red (green) vertical line indicates the corresponding value of the speed of sound for a massive object with $M=2.59\;M_\odot$.}
	\label{fig:Mmaxvs}
\end{figure}

From the observation of Fig.~\ref{fig:MR} a trend across the maximum masses contained in each branch of EoSs, characterized by the speed of sound bound, seems to be inherent. Therefore, we constructed the appropriate diagram of Fig.~\ref{fig:Mmaxvs}. The cross (star) marks represent the maximum masses of $n_{\mathrm{tr}}=1.5n_s$ ($n_{\mathrm{tr}}=2n_s$) case. As the speed of sound bound is getting higher, the marks' color lightens. Similarly to the previous figure, the purple solid horizontal line, with the shaded region, indicates the estimation of the recently observed massive compact object of Ref.~\cite{Abbott-4}. The red (green) curves represent the following fit formula for the $n_{\mathrm{tr}}=1.5n_s$ ($n_{\mathrm{tr}}=2n_s$) case, given below

\begin{equation}
M_{max}=c_1d^{c_2}+c_3d+c_4,
\label{fiteq1}
\end{equation}
where $d=(v_s/c)^2$. The coefficients are given in Table~\ref{tab:table1}.

\begin{table}[H]
	\caption{Parameters of the Eq.~\eqref{fiteq1} and bounds of speed of sound value of Fig.~\ref{fig:Mmaxvs}. The parameters $c_1$, $c_3$, and $c_4$ are in solar mass units $M_\odot$.}
	\centering
	\begin{tabular}{cccccccc}
		\toprule
		\textbf{$n_{\mathrm{tr}}$}	& \textbf{$c_{1}$}	& \textbf{$c_{2}$}		& \textbf{$c_{3}$}		& \textbf{$c_{4}$} 	& \textbf{$(v_s/c)^2_{min}$}		& \textbf{$(v_s/c)^2$}		& \textbf{$(v_s/c)^2_{max}$}\\
		\midrule
		$1.5n_s$ & $-1.6033\times10^{3}$ &$-7.56\times10^{-4}$ & $-1.64\times10^{-1}$ & $1.6068\times10^{3}$ & 0.448 & 0.485 &  0.52 \\
		$2n_s$ & 5.5754 & 0.2742 & -0.6912 & -1.9280  & 0.597 &  0.659  &  0.72\\
		\bottomrule
	\end{tabular}
	\label{tab:table1}
\end{table}

\begin{figure}[h!]
	\centering
	\includegraphics[width=0.9\textwidth]{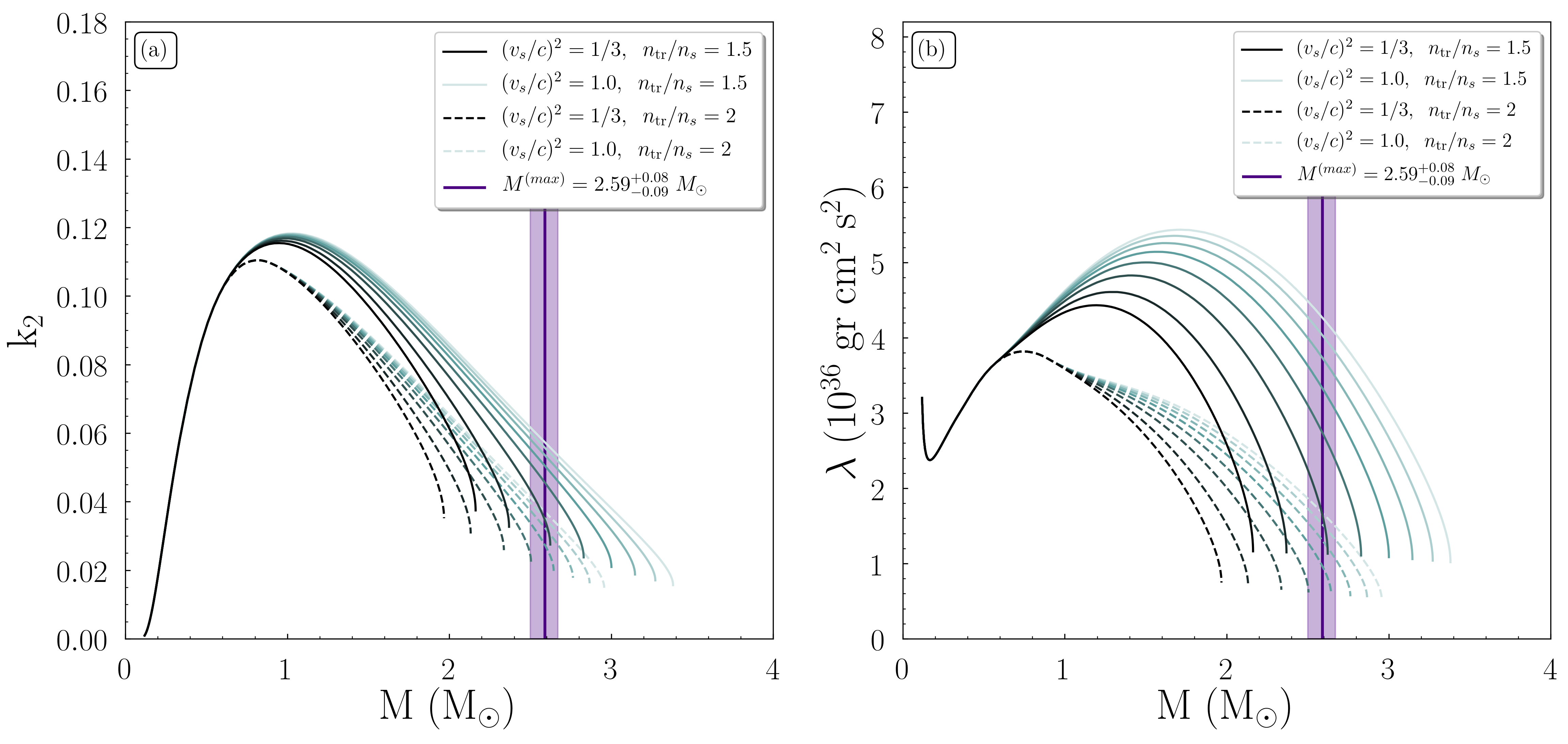}
	\caption{Tidal parameters (a) $k_2$ and (b) $\lambda$ as a function of a NS's mass. The purple vertical line and shaded region indicate the estimation of the recently observed massive compact object of Ref.~\cite{Abbott-4}. The solid (dashed) curves correspond to the $n_{\mathrm{tr}}=1.5n_s$ ($n_{\mathrm{tr}}=2n_s$) case. As the speed of sound is getting higher values, the curves' color lightens.}
	\label{fig:tidalparams}
\end{figure}

By using the mass estimation of the secondary component of GW190814 system, in combination with the fitting formulas mentioned above, we obtained estimations on the speed of sound values for each transition density $n_{\mathrm{tr}}$ scenario. In particular, for a non-rotating massive NS with $M=2.59\;M_\odot$ the value of the speed of sound must be (a)  $(v_s/c)^2 = 0.485$ ($n_{\mathrm{tr}}=1.5n_s$), and (b) $(v_s/c)^2 = 0.659$ ($n_{\mathrm{tr}}=2n_s$). The exact values' interval is given in Table~\ref{tab:table1}. We observe that for higher values of transition density $n_{\mathrm{tr}}$ the fitted curve and marks are shifted downwards; hence the higher the point of the transition in density, the smaller the provided maximum mass. The higher values of speed of sound are more suitable to describe such massive NSs, until a specific boundary value of transition density $n_{\mathrm{tr}}$ in which even the causality would not be suitable. Therefore, a very massive non-rotating NS favors higher values of speed of sound than the $v_s=c/\sqrt{3}$ limit. We notice that  a lower bound on the transition density $n_{\mathrm{tr}}$ is needed to be able in the description of the observed NS mergers~\cite{Kanakis-2020}. Therefore there is a contradiction since the transition density $n_{\mathrm{tr}}$ must be above a specific lower limit and not big enough to predict very massive masses. This kind of remark arises in the speed of sound value, respectively. 
 
Moving on to the tidal parameters, we investigated the tidal Love number $k_2$ and the tidal deformability $\lambda$. In Fig.~\ref{fig:tidalparams} we display the two tidal parameters for the single NS case that we examined. In both diagrams, the vertical purple shaded region and line correspond to the GW190814 system's secondary component compact object. There are two main families of EoSs, distinguished by the transition density $n_{\mathrm{tr}}$. In general, the EoSs with higher values of speed of sound bounds lead to larger values on both tidal parameters. Therefore, a NS with a higher speed of sound more easily deformable, rather than a more compact star (smaller tidal deformation) with lower speed of sound. As Fig.~\ref{fig:tidalparams} shows, the EoSs with smaller transition density $n_{\mathrm{tr}}$ and higher $(v_s/c)^2$ values are more likely to predict a very massive NS of $M=2.59\;M_\odot$. We postulate that a further study with higher transition density $n_{\mathrm{tr}}$ would lead to smaller values of tidal parameters, therefore to more compact stars and more difficult to be deformed. In this case, a very high value of speed of sound, even close to the causal limit, would be necessary to predict such a massive non-rotating NS.

\subsubsection{A very massive neutron star component}
Regarding the binary NS system case, we considered the scenario of a very heavy component NS, in agreement with the recent observation of GW190814 event~\cite{Abbott-4}. Especially, we consider a heavy mass of $m_1=2.59\; M_\odot$ and we let the second star to fluctuate within the range $m_2\in(1,2.59)\;M_\odot$. By subtracting the component masses $m_1, m_2$ in  Eq.~(\ref{chirpmass}) we obtain the corresponding range for the values of $\mathcal{M}_c$. Since the masses are defined, from the  Eqs.~(\ref{L-tild-1}) and~(\ref{Lamb-1}), the effective tidal deformability $\tilde{\Lambda}$ can be determined.

\begin{figure}[h!]
\centering
\includegraphics[width=0.9\textwidth]{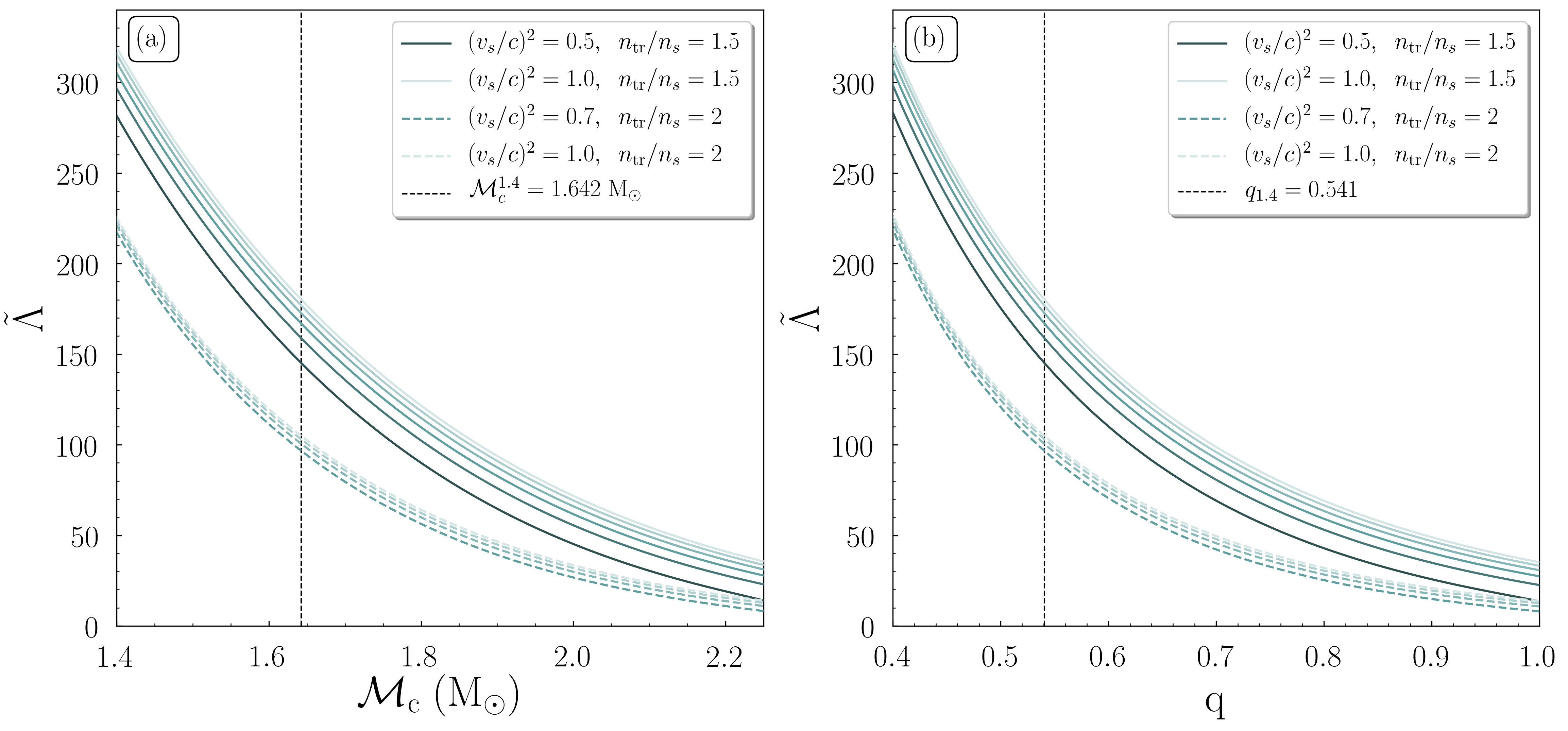}
\caption{The effective tidal deformability $\tilde{\Lambda}$ as a function of (a) the chirp mass $\mathcal{M}_c$ and (b) binary mass ratio $q$, in the case of a very massive NS component, identical to Ref.~\cite{Abbott-4}. As the speed of sound bound is getting higher, the color of EoSs lightens. The black dashed vertical line indicates (a) the corresponding chirp mass $\mathcal{M}_c$ and (b) mass ratio $q$, of a binary NS system with $m_1=2.59\;M_\odot$ and $m_2=1.4\;M_\odot$ respectively.}
	\label{fig:chirpmassgw190814}
\end{figure}

Fig.~\ref{fig:chirpmassgw190814}(a) shows the effective tidal deformability $\tilde{\Lambda}$ as a function of the chirp mass $\mathcal{M}_c$, for all the possible binary NS systems with such a massive NS component. We notice that from the total sum of EoSs that we studied in the single NS case above, in the binary case we used only those who can predict a NS with $2.59\;M_\odot$ mass. In general, there are two families of EoSs, distinguished by the transition density $n_{\mathrm{tr}}$. Inside each family of EoSs, the EoSs with higher speed of sound value provide higher values of $\tilde{\Lambda}$. We notice that for a binary system with $m_1=2.59\;M_\odot$ and $m_2=1.4\;M_\odot$ the chirp mass is $\mathcal{M}_c=1.642\;M_\odot$. In addition, binary NS systems with both heavy components, therefore higher $\mathcal{M}_c$,  lead to smaller values of $\tilde{\Lambda}$. In such case, possible limits on the lower bound of $\tilde{\Lambda}$ may be more suitable to extract constraints on the EoS. 

In the same way, Fig.~\ref{fig:chirpmassgw190814}(b) shows the dependence of $\tilde{\Lambda}$ to the corresponding binary mass ratio $q$. We notice that this kind of $\tilde{\Lambda}-q$ diagram is different from the usual ones (see in comparison Fig.3 of Ref~\cite{Kanakis-2020}) because the chirp mass $\mathcal{M}_c$ has not a unique value. To be more specific, $\mathcal{M}_c$ is a variable and each point of Fig.~\ref{fig:chirpmassgw190814}(b) corresponds to a different binary NS system with the heavier component in all cases to be a very massive NS of $2.59\;M_\odot$. One can observe a similar behavior of the curves, in analogue to Fig.~\ref{fig:chirpmassgw190814}(a); two main families and the EoSs with higher speed of sound provide higher values of $\tilde{\Lambda}$. As the binary NS systems are more symmetric ($q\rightarrow1$), the binary tidal deformability is getting smaller. The highest values of $\tilde{\Lambda}$ correspond to the most asymmetric binary NS systems. We notice that for a binary system with $m_1=2.59\;M_\odot$ and $m_2=1.4\;M_\odot$ the asymmetry ratio is $q=0.541$.

\begin{figure}[h!]
	\centering
	\includegraphics[width=0.5\textwidth]{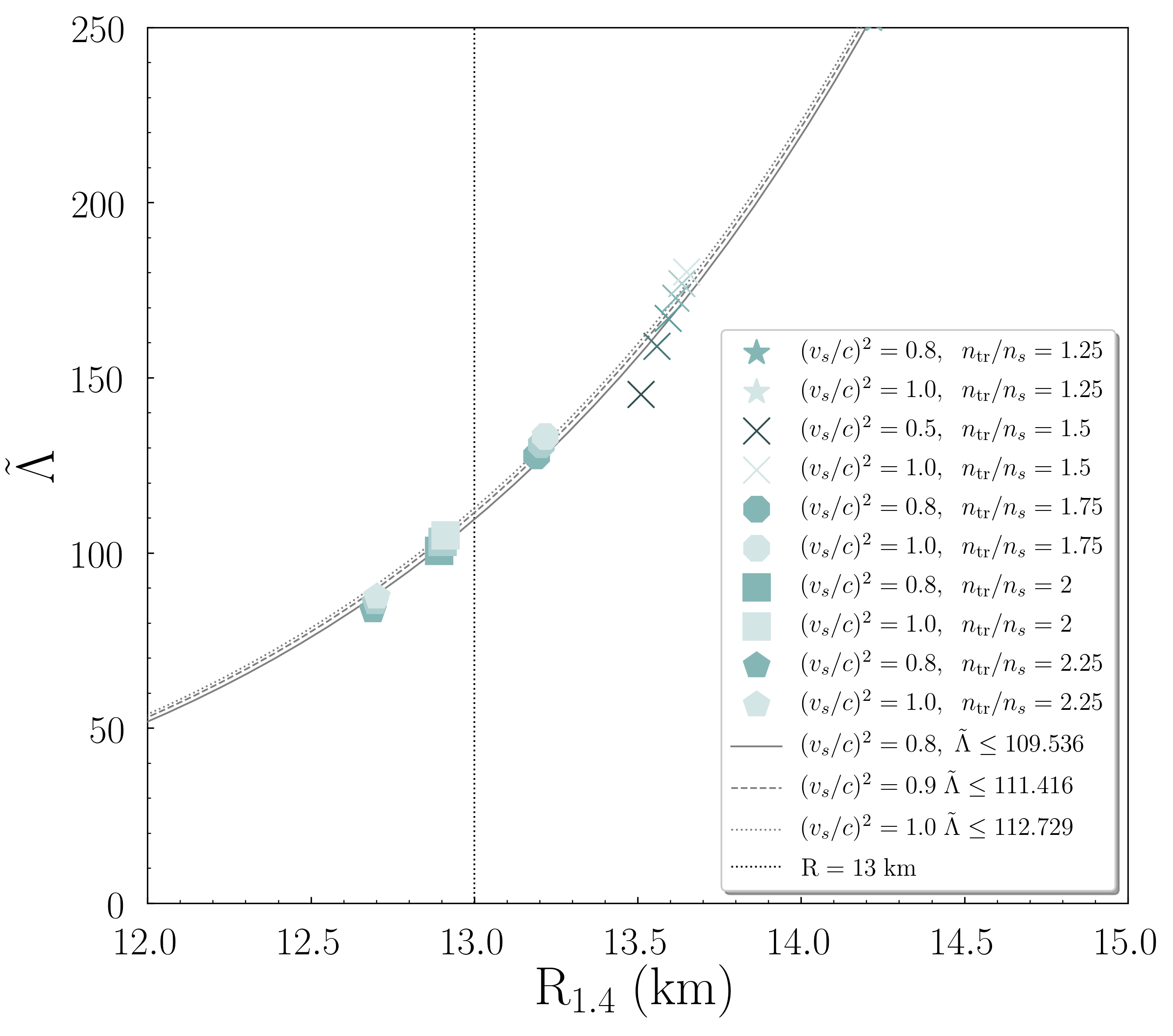}
	\caption{Effective tidal deformability $\tilde{\Lambda}$ vs. radius $R_{1.4}$ of a $m_2=1.4\;M_\odot$ NS. The heavier component of the system was taken to be $m_1=2.59\;M_\odot$. The lighter colors correspond to higher values of speed of sound bounds. The grey lines indicate the expression of Eq.~(\ref{fitR14}). The black dotted vertical line shows the proposed upper limit of Ref.~\cite{Raithel-2018}.}
	\label{fig:LtildeR14}
\end{figure}

Beyond the general behavior of $\tilde{\Lambda}$ that we studied above, it is in our interest to examine the radius and possible constraints that can be derived from it. Following the previous steps, we focus on the $R_{1.4}$ case of a $m_2=1.4\;M_\odot$ secondary component NS, as these values can be extracted from Fig.~\ref{fig:MR}. The heavier component NS is taken to be $m_1=2.59\;M_\odot$. By combining these values with  Fig.~\ref{fig:tidalparams}(b) and   Eqs.~(\ref{L-tild-1}) and~(\ref{Lamb-1}), we obtained the $\tilde{\Lambda}$. In Fig.~\ref{fig:LtildeR14} we display this dependence; the EoSs are in five main groups, characterized by the transition density $n_{\mathrm{tr}}$. We notice that we expanded our study to transition densities $n_{\mathrm{tr}}=[1.25,1.75,2.25]$ to be more accurate in calculations and study in more detail the curves' behavior. The higher speed of sound values correspond to lighter marks' color. In analog to the remarks of the previous Fig.~\ref{fig:chirpmassgw190814}, the high speed of sound bounds lead to higher $\tilde{\Lambda}$ and $R_{1.4}$. Moreover, we applied a fitting expression to the $(v_s/c)^2=[0.8,0.9,1]$ cases. The expression was taken to be in the kind of the proposed relations of Refs.~\cite{Zhao-2018,SoumiDe-2018}.
\begin{equation}
\tilde{\Lambda}=c_5R^{c_6}_{1.4},
\label{fitR14}
\end{equation}
where the coefficients for each case are given in Table~\ref{tab:table2}. A recent study suggested a similar power-law relation that connects the tidal deformability of a single NS to the $R_{1.4}$~\cite{Tsang-2020}. The significance of the tidal deformability $\Lambda_{1.4}$ and $R_{1.4}$ in order to extract information about microscopic quantities was studied in Ref.~\cite{Wei-2020}.

By applying an upper limit on $R_{1.4}$ one can obtain an upper limit on $\tilde{\Lambda}$ for each case. We adopted the general limit of Ref.~\cite{Raithel-2018} that led us to the constraints of Table~\ref{tab:table2}

\begin{table}[H]
\caption{Parameters of the Eq.~\eqref{fitR14} and bounds of $\tilde{\Lambda}$ of Fig.~\ref{fig:LtildeR14}.}
\centering
\begin{tabular}{cccc}
\toprule
\textbf{$(v_s/c)^2$}	& \textbf{$c_{5}$ ($\mathrm{km^{-1}}$)}	& \textbf{$c_{6}$}	& \textbf{$\tilde{\Lambda}$}	\\
\midrule
$0.8$ & $4.1897\times10^{-9}$ &$9.3518$  &$109.536$ \\
$0.9$ & $5.3213\times10^{-9}$ &$9.2652$ &$111.416$  \\
$1$ & $6.1109\times10^{-9}$ &$9.2159$  &$112.729$ \\
\bottomrule
\end{tabular}
\label{tab:table2}
\end{table}

\section{Concluding remarks} \label{sec:remarks}
The GW190814 puzzle and its nature through the nuclear EoS has been addressed in this study. In particular, an effort to explain the existence of a $\sim 2.6~M_{\odot}$ NS, which falls into the NS - black hole mass gap, had been made both for non-rotating and maximally-rotating NSs. For this reason, the MDI-APR EoS and its parametrization for various values of sound speed and transition density in the ranges $(v_{s}/c)^{2}=[1/3,1]$ and $n_{\rm tr} = [1.5,5]~n_{s}$, respectively, have been studied.

Firstly, we compare the MDI-APR EoS with the applicable range of the proposed kerr parameter from Ref.~\cite{Most-2020}, where the authors restraint it using an analytical relation connecting this property with the maximum mass of a non-rotating NS. The results shown that the MDI-APR EoS lies in the range of the gravitational mass of the low mass component, as well as the one for the kerr parameter. In fact, the MDI-APR EoS proposes that the $\sim 2.6~M_{\odot}$ compact star is a rapidly rotating NS, close to its mass-shedding limit.

By considering that the $\sim 2.6~M_{\odot}$ compact star had been rotating at its mass-shedding limit, possible constraints can be extracted for the corresponding equatorial radius. The kerr parameter at the mass-shedding limit can be expressed as a relation with the gravitational mass, and hereafter a region of $\mathcal{K}_{\rm max} = [0.67,0.69]$ is extracted. This region also includes the upper limit of the relevant region from Ref.~\cite{Most-2020} in a narrow range. In addition, using a relation that connects the kerr parameter and the compactness parameter at its mass-shedding limit, a possible tight region for the equatorial radius of the star is implied as $R_{\rm max} = [14.77,14.87]~{\rm km}$.

The upper limit on the central energy/baryon density is a very interesting property, as it is connected with the evolution of the NS and the possible appearance of a phase transition. From this analysis, the central energy density must be lower than the values in the range $\varepsilon_{\rm c}/c^{2} = [2.53,2.89]~10^{15}~{\rm gr}~{\rm cm^{-3}}$, while for the central baryon density, the corresponding range is $n_{c} = [7.27,8.09]~n_{s}$. The latter can inform us about the stability of the NS, as a NS with higher values of central energy/baryon density cannot exist, as well as the appearance of the back-bending process.

The transition density, along with the speed of sound, can infer various structures for the EoS. Assuming that the NSs is a non-rotating one, the transition density is constrained in the region $n_{\rm tr} = [1.5,3.2]~n_{\rm s}$ while the corresponding value of sound speed must be in the range $(v_{s}/c)^{2} = [0.45,1]$. Otherwise, if the NS is considered as a maximally-rotating one, although the speed of sound implies no constraints, the transition density must be higher than $1.6~n_{s}$. 

In the case of non-rotating NS, the construction of the M-R diagram showed  at first glance the cases that can describe the extreme scenario of our study. Moving to a more detailed diagram of the mass vs. the speed of sound bounds, it was feasible to extract stringent constraints on the speed of sound bounds for each case of transition density $n_{\mathrm{tr}}$. For $n_{\mathrm{tr}}=1.5n_s$ this bound is $(v_s/c)^2\in[0.448,0.52]$ while for $n_{\mathrm{tr}}=2n_s$ is $(v_s/c)^2\in[0.597,0.72]$. We observe that the first lower bound is in agreement with the bound extracted above, which is a good validation of our result. We postulate that for higher transition densities $n_{\mathrm{tr}}$ it is more difficult to achieve such a massive non-rotating NS. As the transition density $n_{\mathrm{tr}}$ grows the speed of sound would need to be even close to the causal limit. 

The study of the tidal parameters for a single non-rotating NS allowed us to examine the behavior of EoSs in each case. This lead to the general conclusion that the lower transition densities $n_{\mathrm{tr}}$ lead to higher tidal parameters. Therefore the transition density $n_{\mathrm{tr}}=2n_s$ corresponds to a more compact and less deformable NS. Among the same kind of transition density $n_{\mathrm{tr}}$, the EoSs with higher speed of sound values provide higher tidal parameters. Hence, in a second level across the same kind of $n_{\mathrm{tr}}$ EoSs, the higher speed of sound bound signifies that the tidal deformation is higher and the star is less compact. 

Concerning the binary NS system case, the adoption of a very massive component with $m_1=2.59\;M_\odot$ allowed us to investigate a variety of possible binary NS systems with such a heavy component. We notice that as the binary NS system consists of both heavy component stars, therefore high chirp mass $\mathcal{M}_c$, the effective tidal deformability $\tilde{\Lambda}$ is taking smaller values. Hence, the binary deformation is smaller in such systems. Similarly, the same behavior was noticed in the $\tilde{\Lambda}-q$ diagram in which the increasing binary mass symmetric ratio $q$ leads to smaller values of $\tilde{\Lambda}$. In the case that the second component has a mass $m_2=1.4\;M_\odot$, the chirp mass of the system $\mathcal{M}_c$ and the ratio $q$ are $\mathcal{M}_c=1.642\;M_\odot$ and $q=0.541$ respectively. 

Lastly, we considered the case of a binary NS system with $m_1=2.59\;M_\odot$ with a secondary component $m_2=1.4\;M_\odot$. This selection permitted the study of the radius $R_{1.4}$ and the extraction of possible constraints. In general, the transition density $n_{\mathrm{tr}}=1.5n_s$ provides higher values of $R_{1.4}$ and $\tilde{\Lambda}$ than the $n_{\mathrm{tr}}=2n_s$ case. The examination of other transition densities $n_{\mathrm{tr}}$ permitted us to confirm this behavior. In addition, the high values of speed of sound $(v_s/c)^2$ exhibits a similar behavior; high speed of sound bounds provide higher values on both referred parameters. The adoption of an upper limit on the radius, allowed us to extract some upper limits on $\tilde{\Lambda}$ for each case of sound speed. Nevertheless the value of $n_{\mathrm{tr}}$, as the speed of sound bound is getting higher the upper limit on $\tilde{\Lambda}$ performs an analogous course. We conclude that the existence of such a massive non-rotating NS would require a significant differentiation from all the so far known cases, consisting in any case a unique and very interesting challenge for physics.

\section{Materials and Methods} \label{sec:methods}
The numerical integration of the equilibrium equations for NSs is being under the publicly available RNS code~\cite{Stergioulas-1996} by Stergioulas and Friedman~\cite{Stergioulas-1995}. This code is developed based on the method of Komatsu, Eriguchi, and Hachisu (KEH)~\cite{Komatsu-1989}, while modifications where introduced by Cook, Shapiro, and Teukolsky~\cite{Cook-2-1994}. The input of the code is the EoS in a tabulated form which includes the energy density, the pressure, the enthalpy, and the baryon density.

\authorcontributions{Conceptualization, A.K.P., P.S.K., and C.C.M.; Methodology, A.K.P. and P.S.K.; Software, A.K.P. and P.S.K.; Validation, A.K.P., P.S.K. and C.C.M.; Formal Analysis, A.K.P. and P.S.K.; Investigation, A.K.P., P.S.K., and C.C.M.; Data Curation, A.K.P. and P.S.K.; Writing - Original Draft Preparation, A.K.P., P.S.K., and C.C.M.; Writing - Review $\&$ Editing, A.K.P., P.S.K., and C.C.M.; Visualization, A.K.P., P.S.K., and C.C.M.; Supervision, C.C.M. All authors contributed equally to this work. All authors have read and agreed to the published version of the manuscript.}

\funding{This research received no external funding.}

\acknowledgments{The authors thank  Prof. K. Kokkotas for his constructive comments on the preparation of the manuscript and also Prof. L. Rezzolla for useful correspondence and clarifications.}

\conflictsofinterest{The authors declare no conflict of interest.} 

\abbreviations{The following abbreviations are used in this manuscript:\\

\noindent 
\begin{tabular}{@{}ll}
EoS & Equation of state\\
NS & Neutron star\\
QCD & Quantum chromodynamics\\
MDI & Momentum dependent interaction\\
APR & Akmal, Pandharipande and Ravenhall\\
SNM & Symmetric Nuclear Matter\\
N.R. & Non-rotating configuration\\
M.R. & Maximally-rotating configuration
\end{tabular}}

\reftitle{References}


\begin{thebibliography}{999}
	
	\bibitem{Abbott-4} Abbott, B.P.; et al. GW190814: Gravitational Waves from the Coalescence of a 23 Solar Mass Black Hole with a 2.6 Solar Mass Compact Object. \emph{Astrophys. J. Lett.} {\bf 2020}, \emph{896}, L44.
	
	\bibitem{Datta-2020} Datta, S; Phukon, K.S.; Bose, S. Recognizing black holes in gravitational-wave observations: Telling apart impostors in mass-gap binaries. \emph{arXiv} \textbf{2020}, arXiv:gr-qc/2004.05974.
	
	\bibitem{Alsing-2018} Alsing, J; Silva, H.O.; Berti, E. Evidence for a maximum mass cut-off in the neutron star mass distribution and constraints on the equation of state. \emph{Mon. Not. R. Astron. Soc.} \textbf{2018}, \emph{478}, 1377-1391.
	
	\bibitem{Farr-2020} Farr, W.M.; Chatziioannou, K. A Population-Informed Mass Estimate for Pulsar J0740+6620. \emph{Res. Notes AAS} \textbf{2020}, \emph{4}, 65.
	
	\bibitem{Abbott-2018} Abbott, B.P.; et al. (The LIGO Scientific Collaboration and the Virgo Collaboration) GW170817: Measurements of Neutron Star Radii and Equation of State. \emph{Phys. Rev. Lett.} \textbf{2018}, \emph{121}, 161101.

	\bibitem{Tsokaros-20} Tsokaros, A.; Ruiz, M.; Shapiro, S.L. GW190814: Spin and equation of state of a neutron star companion. \emph{Astrophys. J.} {\bf 2020}, \emph{905}, 48.

	\bibitem{Huang-2020} Huang, K.; Hu, J.; Zhang, Y.; Shen, H. The Possibility of the Secondary Object in GW190814 as a Neutron Star. \emph{Astrophys. J.} {\bf 2020}, \emph{904}, 39.

	\bibitem{Kalogera-2020} Zevin, M.; Spera, M.; Berry, C.; Kalogera, V. Exploring the Lower Mass Gap and Unequal Mass Regime in Compact Binary Evolution. \emph{Astrophys. L. Lett.} {\bf 2020}, \emph{899}, L1.
	
	\bibitem{Fattoyev-2020} Fattoyev, F.J.; Horowitz, C.J.; Piekarewicz, J.; Reed, B. GW190814: Impact of a 2.6 solar mass neutron star on nucleonic equations of state. \emph{Phys. Rev. C} \textbf{2020}, \emph{102}, 065805.

	\bibitem{Essick-2020} Essick, R.; Landry, P.; Discriminating between Neutron Stars and Black Holes with Imperfect Knowledge of the Maximum Neutron Star Mass. \emph{Astrophys. J.} {\bf 2020}, \emph{904}, 80.

	\bibitem{Safarzadeh-2020} Safarzadeh, M.; Loeb, A. Formation of Mass Gap Objects in Highly Asymmetric Mergers. \emph{Astrophys. J. Lett.} {\bf 2020}, \emph{899}, L15. 
	
	\bibitem{Godzieba-2020} Godzieba, D.A.; Radice, D.; Bernuzzi, S. On the maximum mass of neutron stars and GW190814. \emph{arXiv} \textbf{2020}, arXiv:astro-ph.HE/2007.10999.

	\bibitem{Sedrakian-2020} Sedrakian, A.; Weber, F.; Li, J.J. Confronting GW190814 with hyperonization in dense matter and hypernuclear compact stars. \emph{Phys. Rev. D}  {\bf 2020}, \emph{102}, 041301.
	
	\bibitem{Li-2020} Li, J.J.; Sedrakian, A; Weber, F. Rapidly rotating $\Delta$-resonance-admixed hypernuclear compact stars. \emph{Phys. Lett. B} {\bf 2020}, \emph{10}, 135812. 

	\bibitem{Biswas-2020} Biswas, B.; Nandi, R.; Bose, S.; Stergioulas, N. GW190814: On the properties of the secondary component of the binary. \emph{arXiv} \textbf{2020}, arXiv:astro-ph.HE/2010.02090.

	\bibitem{Zhang-2-2020} Zhang, N.B.; Li, B.A. GW190814's Secondary Component with Mass 2.50–2.67 M$_{\odot}$ as a Superfast Pulsar. \emph{Astrophys. J.} {\bf 2020}, \emph{902}, 38.

	\bibitem{Most-2020} Most, E.R.; Papenfort, L.J.; Weih, L.R.; Rezzolla, L. A lower bound on the maximum mass if the secondary in GW190814 was once a rapidly spinning neutron star. \emph{Mon. Not. R. Astron. Soc.} {\bf 2020}, \emph{499}, L82-L86.
	
	\bibitem{Tan-2020} Tan, H.; Hostler-Noronha, J.; Yunes, N. Neutron Star Equation of State in light of GW190814. \emph{Phys. Rev. Lett.} \textbf{2020}, \emph{125}, 261104.
	
	\bibitem{Zhang-2020} Zhang, C.; Mann, R.B. Unified Interacting Quark Matter and its Astrophysical Implications. \emph{arXiv} \textbf{2020}, arXiv:astro-ph.HE/2009.07182.

	\bibitem{Bombaci-2020} Bombaci, I.; Drago, A.; Logoteta, D.; Pagliara, G.; Vidana, I. Was GW190814 a black hole -- strange quark star system? \emph{arXiv} \textbf{2020} arXiv:nucl-th/2010.01509.
	
	\bibitem{Demircik-2020} Demircik, T.; Ecker, C.; Jarvinen, M. Rapidly Spinning Compact Stars with Deconfinement Phase Transition. \emph{arXiv} \textbf{2020}, arXiv:astro-ph.HE/2009.10731. 
	
	\bibitem{Cao-2020} Cao, Z.; Chen, L.W.; Chu, P.C.; Zhou, Y. GW190814: Circumstantial Evidence for Up-Down Quark Star. \emph{arXiv} \textbf{2020}, arXiv:astro-ph.HE/2009.00942.
	
	\bibitem{Dexheimer-2020} Dexheimer, V.; Gomes, R.O.; Klahn, T.; Han. S.; Salinas, M. GW190814 as a massive rapidly-rotating neutron star with exotic degrees of freedom. \emph{arXiv} \textbf{2020}, arXiv:astro-ph.HE/2007.08493. 

	\bibitem{Roupas-2020} Roupas, Z.; Panotopoulos, G.; Lopes, I. QCD color superconductivity in compact stars: color-flavor locked quark star candidate for the gravitational-wave signal GW190814. \emph{arXiv} \textbf{2020}, arXiv:astro-ph.HE/2010.11020.
	
	\bibitem{Rather-2020} Rather, I.A.; Usmani, A.A; Patra, S.K. Hadron-Quark phase transition in the context of GW190814. \emph{arXiv} \textbf{2020}, arXiv:nucl-th/2011.14077.
	
	\bibitem{Moffat-2020} Moffat, J.W. Modified Gravity (MOG) and Heavy Neutron Star in Mass Gap. \emph{arXiv} \textbf{2020}, arXiv:gr-qc/2008.04404.
	
	\bibitem{Oikonomou-2020} Astashenok, A.V.; Capozziello, S.; Odintsov, S.D.; Oikonomou, V.K. Extended gravity description for the GW190814 supermassive neutron star. \emph{Phys. Lett. B} {\bf 2020}, \emph{811}, 135910.
	
	\bibitem{Nunes-2020} Nunes, R.C.; Coelho, G.G.; Araujo, J.C.N. Weighing massive neutron star with screening gravity: a look on PSR J0740 + 6620 and GW190814 secondary component. \emph{Eur. Phys. J. C} {\bf 2020}, \emph{80}, 1115.
	
	\bibitem{Annala-2020} Annala, E.; Gorda, T.; Kurkela, A.; et al. Evidence for quark-matter cores in massive neutron stars. \emph{Nat. Phys.} \textbf{2020}, \emph{16}, 907–910. 
	
	\bibitem{Bombaci-2017} Bombaci, I. The Hyperon Puzzle in Neutron Stars. \emph{JPS Conf. Proc.} \textbf{2017}, \emph{17}, 101002.
	
	\bibitem{Shapiro-1983} Shapiro, S.L.; Teukolsky, S.A. \emph{Black Holes, White Dwarfs, and Neutron Stars}; John Wiley and Sons, New York, 1983.
	
	\bibitem{Glendenning-2000} Glendenning, N.K. \emph{Compact Stars: Nuclear Physics, Particle Physics, and General Relativity}; Springer, Berlin, 2000.
	
	\bibitem{Haensel-2007} Haensel, P.; Potekhin, A.Y.; Yakovlev, D.G. \emph{Neutron Stars 1: Equation of State and Structure}; Springer-Verlag, New York, 2007.
	
	\bibitem{Weinberg-1972} Weinberg, S. \emph{Gravitational and Cosmology: Principle and Applications of the General Theory of Relativity}; Wiley, New York, 1972.
	
	\bibitem{Zeldovich-1971} Zel’dovich, Ya.B.; Novikov, I.D. \emph{Stars and Relativity}; Dover Publications, INC, Mineapolis New York, 1971.
	
	\bibitem{Koliogiannis-2020} Koliogiannis, P.S.; Moustakidis, C.C. Effects of the equation of state on the bulk properties of maximally rotating neutron stars. \emph{Phys. Rev. C} {\bf 2020}, \emph{101}, 015805.
	
	\bibitem{Prakash-1997} Prakash, M.; Bombaci, I.; Prakash, Manju; Ellis, P.J.; Lattimer, J.M.; Knorren, R. Composition and structure of protoneutron stars. \emph{Phys. Rep.} {\bf 1997}, \emph{280}, 1.
	
	\bibitem{Moustakidis-15} Moustakidis, Ch.C. Effects of the nuclear equation of state on the r-mode instability and evolution of neutron stars. \emph{Phys. Rev. C} {\bf 2015}, \emph{91}, 035804.
	
	\bibitem{Gale-1987} Gale, C.; Bertsch, G.; Das Gupta, S. Heavy-ion collision theory with momentum-dependent interactions. \emph{Phys. Rev. C} {\bf 1987}, \emph{35}, 1666.
	
	\bibitem{Gale-1990} Gale, C.; Welke, G.M.; Prakash, M.; Lee, S.J.; Das Gupta, S. Transverse momenta, nuclear equation of state, and momentum-dependent interactions in heavy-ion collisions. \emph{Phys. Rev. C} {\bf 1990}, \emph{41}, 1545.
	
	\bibitem{Akmal-1998} Akmal, A.; Pandharipande, V.R.; Ravenhall, D.G. Equation of state of nucleon matter and neutron star structure. \emph{Phys. Rev. C} {\bf 1998}, \emph{58}, 1804.
	
	\bibitem{Antoniadis-2013} Antoniadis, J; Freire, P.C.; Wex, N.; Tauris, T.M.; Lynch, R.S.; et al. A Massive Pulsar in a Compact Relativistic Binary. \emph{Science} \textbf{2013}, \emph{340}, 1233232.
	
	\bibitem{Cromartie-2016} Cromartie, H.T.; Fonseca, E.; Ransom, S.M.; et al. Relativistic Shapiro delay measurements of an extremely massive millisecond pulsar. \emph{Nat. Astron.} \textbf{2020}, \emph{4}, 72.
	
	\bibitem{Linares-2018} Linares, M.; Shahbaz, T.; Casares, J. Peering into the Dark Side: Magnesium Lines Establish a Massive Neutron Star in PSR J2215+5135. \emph{Astrophys. J.} \textbf{2018}, \emph{859}, 54.
	
	\bibitem{Stergioulas-1996} Stergioulas, N. \textbf{1996}, http://www.gravity.phys.uwm.edu/rns/.
	
	\bibitem{Breu-2016} Breu, C.; Rezzolla, L. Maximum mass, moment of inertia and compactness of relativistic stars. \emph{Mon. Not. R. Astron. Soc.} {\bf 2016}, \emph{459}, 646.
	
	\bibitem{Margaritis-2020} Margaritis, Ch.; Koliogiannis, P.S.; Moustakidis, Ch.C. Speed of sound constraints on maximally rotating neutron stars. \emph{Phys. Rev. D} {\bf 2020}, \emph{101}, 043023.
	
	\bibitem{Rhoades-1974} Rhoades, C.E.; Ruffini, R. Maximum Mass of a Neutron Star. \emph{Phys. Rev. Lett.} {\bf 1974}, \emph{32}, 324.
	
	\bibitem{Kalogera-1996} Kalogera, V.; Baym, G. The Maximum Mass of a Neutron Star. \emph{Astrophys. J.} {\bf 1996}, \emph{470}, L61.
	
	\bibitem{Koranda-1977} Koranda, S.; Stergioulas, N.; Friedman, J.L.	Upper Limits Set by Causality on the Rotation and Mass of Uniformly Rotating Relativistic Stars. \emph{Astrophys. J.} {\bf 1997}, \emph{488}, 799.
	
	\bibitem{Chamel-2013a} Chamel, N.; Haensel, P.; Zdunik, J.L.; Fantina, A.F On the maximum mass of neutron stars. \emph{Int. J. Mod. Phys. E} {\bf 2013}, \emph{22}, 1330018.
	
	\bibitem{Podkowka-2018} Podkowka, D.M.; Mendes, R.F.P.; Poisson, E. Trace of the energy-momentum tensor and macroscopic properties of neutron star. \emph{Phys. Rev. D} {\bf 2018}, \emph{98}, 064057.
	
	\bibitem{Xia-2019} Xia, C.; Zhu, Z.; Zhou, X.; Li, A. Sound velocity in dense stellar matter with strangeness and compact stars. \emph{arXiv} \textbf{2019}, arXiv:nucl-th/1906.00826.
	
	\bibitem{Feynman-1949} Feynman, R.P.; Metropolis, N.; Teller, E. Equations of State of Elements Based on the Generalized Fermi-Thomas Theory. \emph{Phys. Rev.} \textbf{1949}, \emph{75}, 1561.
	
	\bibitem{Baym-1971} Baym, G.; Pethik, C.; Sutherland, P. The Ground State of Matter at High Densities: Equation of State and Stellar Models. \emph{Astrophys. J.} \textbf{1971}, \emph{170}, 299.
	
	\bibitem{Tews-2018} Tews, I.; Carlson, J.; Gandolfi, S.; Reddy, S. Constraining the Speed of Sound inside Neutron Stars with Chiral Effective Field Theory Interactions and Observations. \emph{Astrophys. J.} \textbf{2018}, \emph{860}, 149.
	
	\bibitem{Postnikov-2010} Postnikov, S.; Prakash, M.; Lattimer, J.M. Tidal Love numbers of neutron and self-bound quark stars. \emph{Phys. Rev. D} \textbf{2010}, \emph{82}, 024016.
	
	\bibitem{Baiotti-2019} Baiotti, L. Gravitational waves from neutron star mergers and their relation to the nuclear equation of state. \emph{Prog. Part. Nucl. Phys.} \textbf{2019}, \emph{109}, 103714.
	
	\bibitem{Flanagan-08} Flanagan, E.E.; Hinderer, T. Constraining neutron-star tidal Love numbers with gravitational-wave detectors. \emph{Phys. Rev. D} \textbf{2008}, \emph{77}, 021502(R).
	
	\bibitem{Hinderer-08} Hinderer, T. Tidal Love Numbers of Neutron Stars. \emph{Astrophys. J.} \textbf{2008}, \emph{677}, 1216.
	
	\bibitem{Damour-09} Damour, T.; Nagar, A. Relativistic tidal properties of neutron stars. \emph{Phys. Rev. D} \textbf{2009}, \emph{80}, 084035.
	
	\bibitem{Hinderer-10} Hinderer, T.; Lackey, B.D.; Lang, R.N.; Read, J.S. Tidal deformability of neutron stars with realistic equations of state and their gravitational wave signatures in binary inspiral. \emph{Phys.Rev. D} \textbf{2010}, \emph{81}, 123016.
	
	\bibitem{Fattoyev-13} Fattoyev, F.J.; Carvajal, J.; Newton, W.G.; Li, B.A. Constraining the high-density behavior of the nuclear symmetry energy with the tidal polarizability of neutron stars.  \emph{Phys. Rev. C} \textbf{2013}, \emph{87}, 015806.
	
	\bibitem{Lackey-015} Lackey, B.D.; Wade, L. Reconstructing the neutron-star equation of state with gravitational-wave detectors from a realistic population of inspiralling binary neutron stars. \emph{Phys. Rev. D} \textbf{2015}, \emph{91}, 043002.
	
	\bibitem{Takatsy-2020} Tak\'atsy, J.; Kov\'acs, P. Comment on “Tidal Love numbers of neutron and self-bound quark stars”. \emph{Phys. Rev. D} \textbf{2020}, \emph{102 (2)}, 028501.
	
	\bibitem{Thorne-1998} Thorne, K.S.; Tidal stabilization of rigidly rotating, fully relativistic neutron stars. \emph{Phys. Rev. D} \textbf{1998}, \emph{58}, 124031.
	
	\bibitem{Abbott-1} Abbott, B.P.; et al. GW170817: Observation of Gravitational Waves from a Binary Neutron Star Inspiral. \emph{Phys. Rev. Lett.} \textbf{2017}, \emph{119}, 161101.

	\bibitem{Abbott-3} Abbott, B.P.; et al. Properties of the Binary Neutron Star Merger GW170817. \emph{Phys. Rev. X} \textbf{2019}, \emph{9}, 011001.
	
	\bibitem{Koliogiannis-2-2020} Koliogiannis, P.S.; Moustakidis, Ch.C. Thermodynamical description of hot rapidly rotating neutron stars and neutron stars merger remnant. \emph{arXiv} \textbf{2020}, arXiv:astro-ph.HE/2007.10424 .
	
	\bibitem{Cook-1994} Cook, G.B.; Shapiro, S.L.; Teukolsky, S.A. Rapidly Rotating Neutron Stars in General Relativity: Realistic Equations of State. \emph{Astrophys. J.} \textbf{1994}, \emph{424}, 823.
	
	\bibitem{Salgado-1994} Salgado, M.; Bonazzola, S.; Gourgoulhon, E.; Haensel, P. High precision rotating neutron star models 1: Analysis of neutron star properties. \emph{Astron. Astrophys.} \textbf{1994}, \emph{291}, 155.
	
	\bibitem{Kanakis-2020} Kanakis-Pegios, A.; Koliogiannis, P.S.; Moustakidis, Ch.C. Speed of sound constraints from tidal deformability of neutron stars. \emph{Phys. Rev. C} \textbf{2020}, \emph{102}, 055801.
	
	\bibitem{Zhao-2018} Zhao, T.; Lattimer, J.M. Tidal deformabilities and neutron star mergers. \emph{Phys. Rev. D} \textbf{2018}, \emph{98}, 063020.
	
	\bibitem{SoumiDe-2018} De, S.; Finstad, D.; Lattimer, J.M.;  Brown, D.A.; Berger, E.; Biwer, C.M. Tidal Deformabilities and Radii of Neutron Stars from the Observation of GW170817. \emph{Phys. Rev. Lett.} \textbf{2018}, \emph{121}, 091102.

	\bibitem{Tsang-2020} Tsang, C.Y.; Tsang, M.B.; Danielewicz, P.; Fattoyev, F.J.; Lynch, W.G. Insights on Skyrme parameters from GW170817. \emph{Phys. Lett. B} {\bf 2019}, \emph{796}, 1.
	
	\bibitem{Wei-2020} Wei, J.-B.; Lu, J.-J.; Burgio, G.F.; Li, Z.-H.; Schulze, H.-J. Are nuclear matter properties correlated to neutron star observables? \emph{Eur. Phys. J. A} {\bf 2020}, \emph{56}, 63.

	\bibitem{Raithel-2018} Raithel, C.A.; Özel, F.; Psaltis, D. Tidal Deformability from GW170817 as a Direct Probe of the Neutron Star Radius. \emph{Astrophys. J. Lett.} \textbf{2018}, \emph{857}, L23.
	
	\bibitem{Stergioulas-1995} Stergioulas, N.; Friedman, J.L. Comparing Models of Rapidly Rotating Relativistic Stars Constructed by Two Numerical Methods. \emph{Astrophys. J.} \textbf{1995}, \emph{444}, 306.
	
	\bibitem{Komatsu-1989} Komatsu, H.; Eriguchi, Y.; Hachisu, I. Rapidly rotating general relativistic stars. I - Numerical method and its application to uniformly rotating polytropes. \emph{Mon. Not. R. Astron. Soc.} \textbf{1989}, \emph{237}, 355.
	
	\bibitem{Cook-2-1994} Cook, G.B.; Shapiro, S.L.; Teukolsky, S.A. Rapidly Rotating Polytropes in General Relativity. \emph{Astrophys. J.} \textbf{1994}, \emph{422}, 227.

\end{thebibliography}
\end{document}